%% file: ms.tex
\documentclass[12pt,preprint]{aastex}
\shorttitle{Abundances on the Main Sequence of Omega Centauri}
\shortauthors{Stanford et al.}
\newcommand{\wcen}{$\omega$ Cen}
\newcommand{\feh}{[Fe/H]}

\newcommand{\eq}{$\approx$}

\newcommand{\tuc}{47~Tuc}

\begin{document}

\title{Abundances on the Main Sequence of Omega Centauri}

\author{Laura M. Stanford, G. S. Da
 Costa and John E. Norris}
\affil{Research School of Astronomy and Astrophysics, Australian
  National University, Weston, ACT, 2611, Australia}
\email{stanford, gdc, jen@mso.anu.edu.au}
\and

\author{Russell D. Cannon} \affil{Anglo-Australian
Observatory, P.O. Box 296, Epping, NSW, 2121, Australia}
\email{rdc@aaoepp.gov.au}

\begin{abstract}
Abundance ratios of carbon, nitrogen and strontium relative to iron,
calculated using spectrum synthesis techniques, are given for a sample
of main sequence and turnoff stars that belong to the globular cluster
$\omega$ Centauri.  The variations of carbon, nitrogen and/or
strontium show several different abundance patterns as a function of
[Fe/H].  The source of the enhancements/depletions in carbon, nitrogen
and/or strontium may be enrichment from asymptotic giant branch stars
of low (1--3 M$_{\odot}$) and intermediate (3--8 M$_\odot$) mass.
Massive rotating stars which produce excess nitrogen without carbon
and oxygen overabundances may also play a role.  These abundances
enable different contributors to be considered and incorporated into
the evolutionary picture of $\omega$ Cen.

\end{abstract}

\keywords{globular clusters: general ---
globular clusters: individual ($\omega$ Centauri)}

\section{Introduction} \label{intro}
\defcitealias{nd95}{ND95}

The globular cluster $\omega$ Centauri is known to have
characteristics not shared by other clusters.  Its unusual nature was
first noted by \citet{woo66} and \citet{cs73}, who demonstrated that
its red giant branch (RGB) has an intrinsic color spread.  Early
abundance studies of member stars, obtained spectroscopically, have
shown a range in metallicity \citep{fr75, le77, bde78, cb86, fss88,
pn89}.  In the last decade detailed spectroscopic and photometric work
on {\wcen} has revealed the large extent of the abundance range within
the system.

The cluster metallicity distribution has been studied in more detail
recently, using spectroscopic analysis of the Ca II triplet lines
\citep{nfm96, sk96}, photometry \citep{lee99, pan00, hr00, sol05a} and
spectroscopic analysis of the Ca II K line \citep{sta06a}.  These
studies found few, if any, stars with {\feh}$<$--2.0.  The
distribution then rises to a peak at {\feh}=--1.7 with a long tail to
higher metallicities, up to {\feh}$\approx$--0.4. The distribution
also has a second peak at {\feh}=--1.2 \citep{nfm96}.  Three main
populations within the cluster have been identified \citep{nfm96,
sk96,pan02}.  The bulk of the cluster stars ($\sim$70\%) has
metallicities at [Fe/H]$\approx$--1.7, the metal-intermediate
population has [Fe/H]$\approx$--1.2 and represents roughly 20\% of the
stars, and the remaining 5\% of the stars are metal-rich at
[Fe/H]$\approx$--0.6.  \citet{sol05a} found the RGB to be discrete in
nature and identified up to five distinct populations, three of which
are within the metal-intermediate population described above.
Elemental studies have shown a range in abundances within the cluster
members.  Carbon, nitrogen and oxygen exhibit large scatters for a
given {\feh} (\citealt{per80, bw93}; \citealt[hereafter ND95]{nd95}).
The $\alpha$-elements (Mg, Si, Ca and Ti) have constant values at a
given [Fe/H] for most metallicities (\citealt{bw93, scl95};
\citetalias{nd95}; \citealt{smi00}) but [$\alpha$/Fe] decreases at
higher metallicities ({\feh} $>$ --1.0) \citep{pan02}.  As the
$\alpha$ element abundances are constant with iron below
{\feh}\eq--1.0, they are produced by the same source, most likely
supernovae Type II.  At higher abundances, decreasing [$\alpha$/Fe] is
indicative of supernovae Type Ia contributions.

The heavy neutron-capture elements (e.g. Y, Zr, Ba, La, Pr, Nd) give
information on enrichment sources, shedding light on whether a slow or
rapid neutron-capture process occurred (or both).  \citetalias{nd95}
and \citet{scl95} found the s-process element abundance ratios with
respect to iron increase with increasing {\feh} and then flatten above
{\feh}=--1.2.  These results are in contrast to normal globular
clusters, and indicate that the stellar winds from asymptotic giant
branch (AGB) stars (sources of s-process elements) were involved with
the enrichment process of {\wcen}. \citetalias{nd95} concluded also
that there were no correlations between the abundance ratios with
respect to iron for the s-process elements and those for C, N, O, Na
and Al.  Eu is an r-process element, and abundances of this element
are found to be lower in {\wcen} stars than in the field
\citep{scl95}, and ratios of Ba/La show that the enrichment process is
dominated by the s-process \citepalias{nd95}.

Another piece in the {\wcen} puzzle is the double main sequence seen
in deep photometry of the cluster \citep{and02, bed04}.  This is
thought to be the result of helium variations of the order of
$\Delta$Y$\sim$0.12~dex between two populations \citep{nor04, pio05}.
The source of the large amount of additional helium required may be
AGB stars with masses larger than 6M$_{\odot}$, stellar winds
associated with early evolutionary phases of massive stars, or perhaps
Type II supernovae \citep{nor04,pio05, mm06}.  \citet{bn06} suggest
that the helium enhanced population formed from gas ejected by the
stellar populations that surrounded {\wcen} when it was the nucleus of
a dwarf galaxy.  \citet{mm06} have argued that in models with low
metallicity, massive stars with high rotation can produce the large
amounts of helium required to explain the blue main sequence (bMS),
depending on the slope of the initial mass function.  These models
also produce a large excess of nitrogen but little carbon, consistent
with the phenomena observed in spectra of the stars in the bMS
\citep{pio05}.

Variations of the CNO group, Na, Mg and Al abundances have been found
in most {\it normal} globular clusters on the RGB to different
degrees.  In the last few decades, studies of main sequence turn off
(MSTO) stars in a number of globular clusters have shown that
variations exist in C and N in this region of the CMD as well.  These
clusters include 47 Tuc \citep{can98, bri04}, NGC~6752 \citep{ss91,
gra01a}, M71 \citep{coh99, bc01, rc02}, M5 \citep{cbs02} and M13
\citep{bcs04}.  Anticorrelations between Na and O have been shown to
exist on the main sequence (MS) in 47 Tuc \citep{car04}, NGC~6397 and
NGC~6752 \citep{gra01b}.  These latter two clusters also have an
anticorrelation between Mg and Al \citep{gra01b}.

There are three possibilities for the origin of the abundance
variations in {\wcen} and other globular clusters: internal
processing, accretion of matter onto surface layers, and primordial
formation.  Firstly, internal processing within the individual stars
themselves, with the products then being mixed to the surface layers,
may be the source of the abundance variations.  This may account, in
part, for the variations in the RGB stars in some clusters, as the
level of C depletion on the RGB increases with decreasing magnitude
(e.g. \citealt{lan86}, M92).  However, mixing of the internal layers
does not occur in the current MS stars (e.g. \citealt{dd82}).
Therefore, internal processing cannot explain all the variations seen
in globular clusters.  The second option is that the enriched objects
accreted matter onto their surface layers from the stellar winds of
AGB stars.  The stars may have been a part of a binary system or in
the vicinity of AGB stars. The accreted material will be on the
surface layers only, and therefore, in a comparison of MS with RGB
abundance ratios, one should generally find higher abundances for the
unevolved stars compared with the evolved ones as increased convective
mixing as the stars evolve up the RGB will dilute the accreted
material.  Primordial variations account for the third and final
possibility.  Here, the stars formed from material that had been
previously enriched from sources such as winds and ejecta of AGB
stars, massive rotating stars and/or supernovae.

Most spectroscopic studies have concentrated on the red giant branch
(RGB) stars in {\wcen} because of their greater brightness.  We have
obtained spectra of a sample of turnoff stars to examine the
relationship between age and metallicity in the cluster
\citep{sta06a}, and to determine abundances of several elements and
compare them with those found on the RGB.  This will give further
insights into the evolutionary history of {\wcen} and enable us to
distinguish between surface enhancement and primordial enrichment
scenarios within the cluster.

The present paper describes the analysis of these spectra to determine
abundances of C, N and strontium, relative to iron, for the MS and
turnoff (TO) sample and an attempt to untangle the complex
evolutionary history of {\wcen}.  Early results of this work are given
in \citet{dac05}. In \S~2 we describe the observations and reduction
techniques. \S~3 outlines the derivation of metallicities and ages for
the sample and discusses the metallicity populations. We describe the
techniques used to determine C, N and Sr abundances in \S~4.  Finally,
\S~5 summarizes the results and discusses a chemical evolutionary
history for the cluster.

\section{Observations and Reduction} \label{OR_sect}

\begin{figure}
\begin{center}
\includegraphics[scale=.60,angle=0]{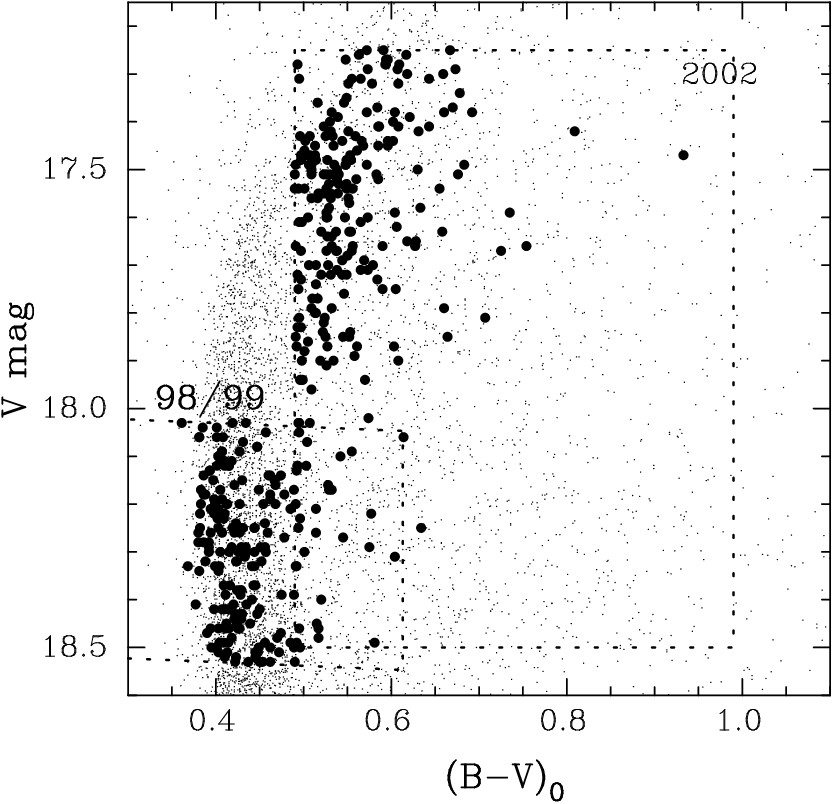}
\figcaption[f1.ps]{Color-magnitude diagram for stars near the MSTO and
subgiant branch of {\wcen}.  Large circles denote stars which are
radial velocity members from two samples selected within the regions
outlined. \label{cmd1} }
\end{center}
\end{figure}

The observations and reduction of our data are described in detail in
\citet{sta06a} to which we refer the reader.  Suffice it here to say
the photometry of the cluster was obtained with the 1m
telescope/Tektronix CCD combination at Siding Spring Observatory in
the V and B bands.  From these data a CMD, shown in Figure \ref{cmd1},
was constructed for objects within an annulus 15--25 arcminutes from
the cluster center. Two regions were defined on the upper MS with a
view to determining the metallicity and age range for the cluster.
Stars in each region were observed using the Two degree Field (2dF)
Multiobject spectrograph \citep{lew02} on the Anglo-Australian
Telescope . This spectrograph has the capability of simultaneously
observing up to 400 objects using a fibre fed system.  The first
sample was observed in May 1998 and April 1999 (hereafter 98/99
sample), and the second in March 2002 (hereafter 2002 sample). Figure
\ref{cmd1} shows the two boxes from which candidates were selected.
Spectra were obtained with 1200 line/mm gratings employed in the
spectrographs and covered the wavelength range
$\lambda\lambda$3800--4600\AA.  They have a resolution of
$\sim$2.5\AA\ FWHM and signal-to-noise (S/N) of 30-50.  A final sample
of 424 radial velocity selected members was found. The members are
shown as large dots in Figure \ref{cmd1}. The small dots represent the
photometry of all the remaining objects that do not have membership
information.

Data for stars on the MS were also obtained for three other globular
clusters --- NGC 6397, NGC 6752 and 47 Tuc.  These clusters were
observed in a similar manner to $\omega$ Cen, and membership was
determined based on radial velocities and metallicities.  The data
obtained for the clusters were used to test the reliability of the
metallicity calibration, and for comparison of abundances.

\section{Metallicities and Populations} \label{met_sect}

The process for calculating metallicities has been described in detail
in \citet{sta06a}.  To summarize, the metallicities were calculated
using a combination of two methods (see \citet{bee99} for details of
these methods briefly described below).  The first uses the Ca
\textsc{ii} K line strength and color of an object to assign a
metallicity.  Due to the saturation of the Ca \textsc{ii} K line these
metallicities become uncertain above [Fe/H]$\approx$--1.0.  The second
method utilizes the weak metal lines in the spectrum.  This method is
reliant on spectra having good signal-to-noise ($>$30) as the noise
can become confused with the weak metal lines.  As metal-poor stars
have few and very weak metal lines, this method is unreliable at lower
metallicities ([Fe/H]$<$--2.0) and the abundance from the Ca
\textsc{ii} K line strength was adopted.  Between [Fe/H]=--2.0 and
[Fe/H]=--1.0 a weighted mean of the two values was used and above
\feh=--1.0 metallicities from the second method were adopted.
Metallicities were calculated for the {\wcen} sample, and for the
calibrating clusters NGC~6397, NGC~6752 and {\tuc}.  An age was then
calculated for each {\wcen} star based on its position on the CMD and
metallicity using theoretical isochrones (cf. \citet{sta06a}).

\begin{figure}
\begin{center}
\includegraphics[scale=.60,angle=0]{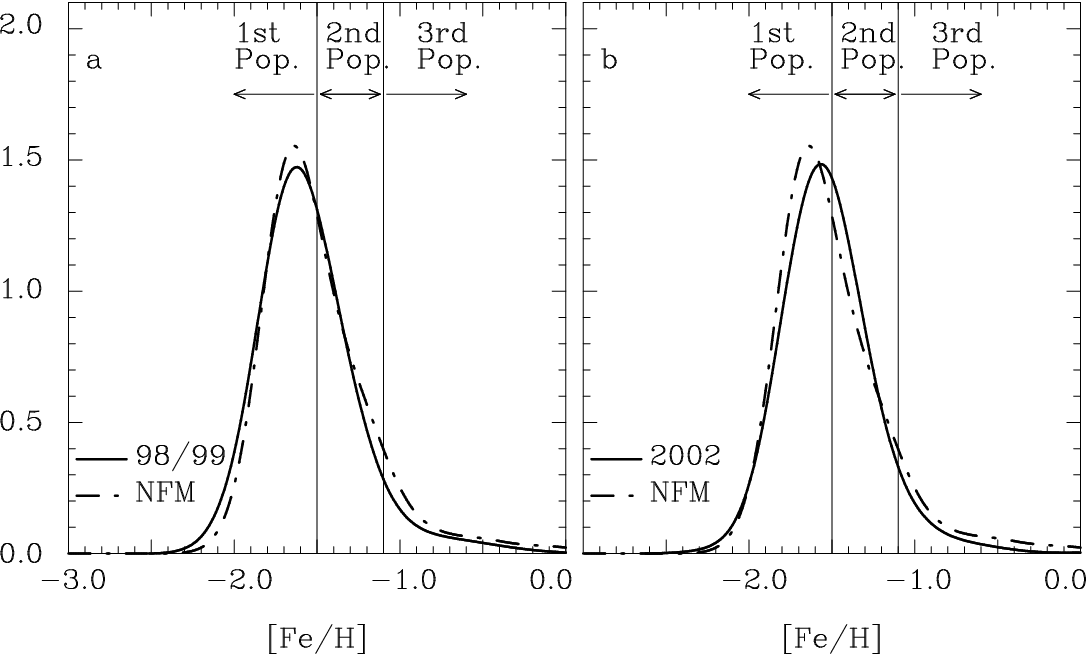}
\figcaption[f2.ps]{Generalized histograms of the metallicity
distribution for the 98/99 (panel~a) and 2002 (panel~b) samples
(gaussian kernel of $\sigma$=0.15--0.20, based on the individual
errors for each metallicity).  For comparison, the \citet{nfm96} red
giant branch distribution is also plotted ($\sigma$=0.14). Indicated
on each histogram is the division into three populations with the
first having [Fe/H]$<$--1.5, the second --1.5$\leq$[Fe/H]$<$--1.1, and
the third [Fe/H]$\geq$--1.1.
\label{hist} }
\end{center}
\end{figure}

The metallicities are used here to divide the samples into three
groups, approximate the three main populations in {\wcen}.  The
metal-poor population (1st Pop.) contains stars with [Fe/H]$<$--1.5.
The metal-intermediate population (2nd Pop.)  contains objects that
have --1.5$\leq$[Fe/H]$<$--1.1.  The third population (3rd Pop.)
contains the stars that have the highest metallicities in our sample
with [Fe/H]$\geq$--1.1.  The histograms of metallicities for our two
samples (98/99 and 2002) are shown in Figure \ref{hist}.  The figure
also shows the cuts in metallicity that make up the three populations,
with 255/424 in the 1st Population, 144/424 in the second and 25/424
in the metal-rich population.  The calibrating clusters were used to
compare abundances at similar metallicities. The values for these
clusters come from \citet{har96}. NGC~6397, with [Fe/H]=--1.96, was
compared with the first population.  NGC~6752 is borderline between
the 1st Pop. and 2nd Pop. with [Fe/H]=--1.56, and can be used for
comparison purposes with both of these populations.  {\tuc}, with
[Fe/H]=--0.76, was used to compare with the 3rd Pop.

\begin{figure}
\begin{center}
\includegraphics[scale=.60,angle=0]{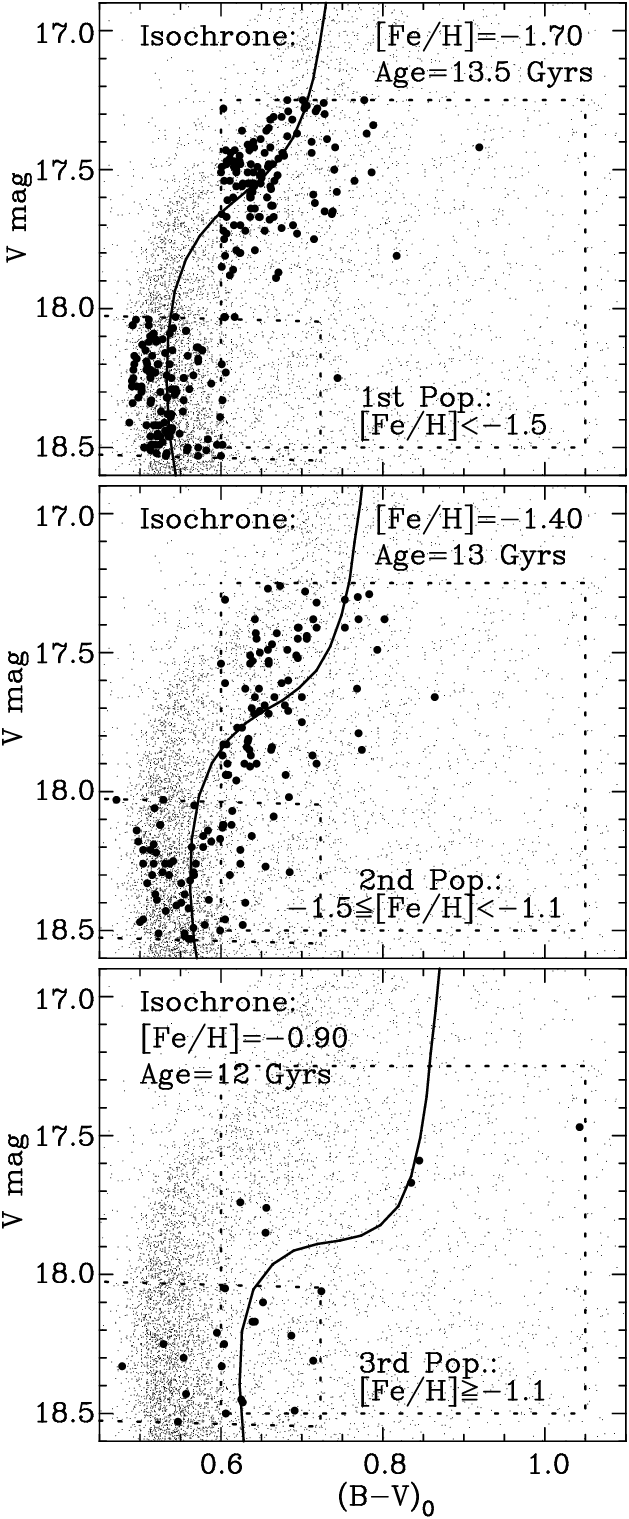}
\figcaption[f3.ps]{Color-magnitude diagrams of the three populations
of {\wcen} defined by their metallicities. Isochrones with
metallicities and ages representing the means of each population are
also plotted.
\label{cmd2} }
\end{center}
\end{figure}

 The photometry for the three {\wcen} populations is plotted in Figure
\ref{cmd2}. An isochrone (Y$^{2}$, \citealt{yi01}) was fitted to each
population on the CMD with parameters depending on the mean
metallicity and age for the population.  These isochrones are
overplotted in Figure \ref{cmd2}.  Ten points were chosen along the
isochrone that encompassed our sample in luminosity and color and the
corresponding temperature and gravity were used in the abundance
analysis.  For the first and second populations (at {\feh}=--1.7
and~--1.2 respectively) the [$\alpha$/Fe] ratio was taken to be 0.3
(\citetalias{nd95}; \citealt{scl95, smi00}).  For the most metal-rich
population with a mean [Fe/H]=--0.8, the alpha enhancement was taken
to be [$\alpha$/Fe]=0.18.

\subsection{Indices}

\begin{figure}
\begin{center}
\includegraphics[scale=.80,angle=0]{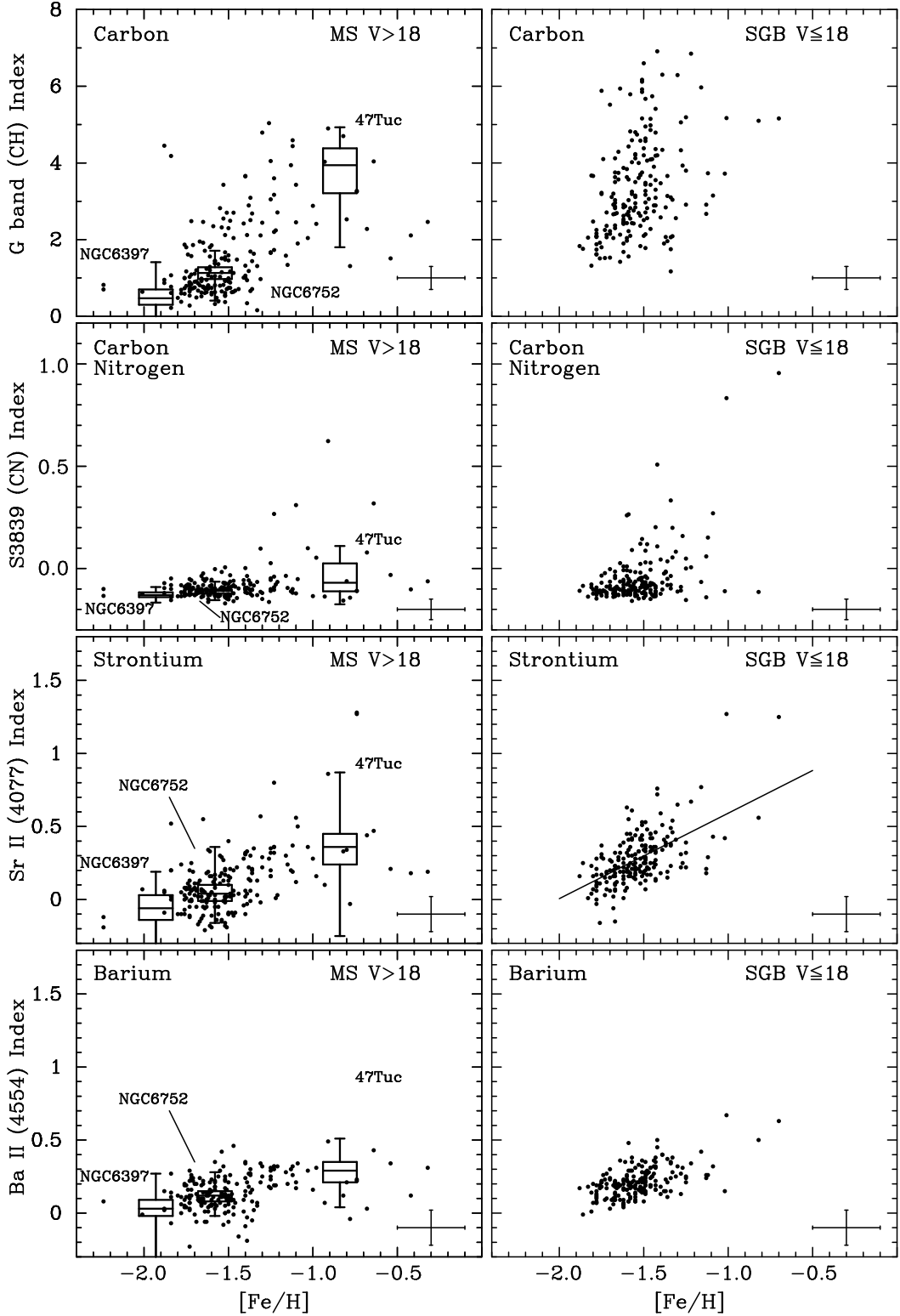} \figcaption[f4.ps]{ Indices
for the G band (CH), CN at $\sim$3883{\AA}, Sr {\sc ii} 4077{\AA} and
Ba {\sc ii} 4554{\AA}, plotted against metallicity for the main
sequence (V$>$18) and subgiant branch stars (V$\leq$18).  The boxed
regions on the MS plot represent indices calculated for three other
globular clusters --- NGC~6397 ([Fe/H]=--1.96), NGC~6752
([Fe/H]=--1.56) and {\tuc} ([Fe/H]=--0.76). \label{indfeh} }
\end{center}
\end{figure}

Indices were measured for the G band (CH) at $\sim$4300{\AA}, violet
CN band at 3883{\AA}, the Sr \textsc{ii} 4077{\AA} and 4215{\AA}
lines, and Ba \textsc{ii} 4554{\AA} for the {\wcen} members and
calibrating cluster stars. The weaker Sr \textsc{ii} 4215{\AA} line
was used as confirmation of the Sr abundance, which was determined
primarily from the Sr\textsc{ii} 4077{\AA} feature.  The bandpasses
for CH, Sr and Ba are given in Table \ref{tbl-1} and the indices are
defined following \citet{bee99}.  The CN index is S3839 as defined by
\citet{nor81}.  The {\wcen} data are plotted as a function of
metallicity in Figure \ref{indfeh}, and are split into main-sequence
(V$>$18) and SGB (V$\leq$18) samples.  The corresponding indices
determined for NGC~6397, NGC~6752 and {\tuc} are represented by
boxplots showing the area in [Fe/H]--index space they occupy for
comparison with the MS {\wcen} stars, as we do not have data for the
SGB stars in these clusters.  The center line within the rectangle for
each boxplot represents the median for each dataset.  The lower and
upper sides of the box represent the first and third quartiles while
the vertical lines extending from the upper and lower edges of the box
represent the upper and lower limits of the data.  The error bars in
the figure represent the standard error of the mean of the indices for
individual observations, and the derived error in metallicity.

Studying the two CH plots, one can see that the SGB group has, in
general, higher indices than on the MS one.  This is due to the cooler
temperatures on the SGB than the MS.  At lower metallicities on the
MS, there are two members that have unusually high C indices.  These
fall well outside the ranges in [Fe/H] and the CH index that is
occupied by the calibrating clusters.  Outliers can also be identified
on the SGB.  The S3839 index shows several objects which appear to
have high abundances of carbon and/or nitrogen in both the MS and SGB
groups.  The strontium index plot shows a large scatter, with several
outliers in each.  For the SGB group, a least-squares fit to the data
was performed, shown by the solid line.  This line indicates there are
several objects with considerably higher Sr indices than the main
population.  There is little spread in the Ba index, most likely due
to the low sensitivity of our data to the Ba abundance.

\begin{figure}
\begin{center}
\includegraphics[scale=.60,angle=0]{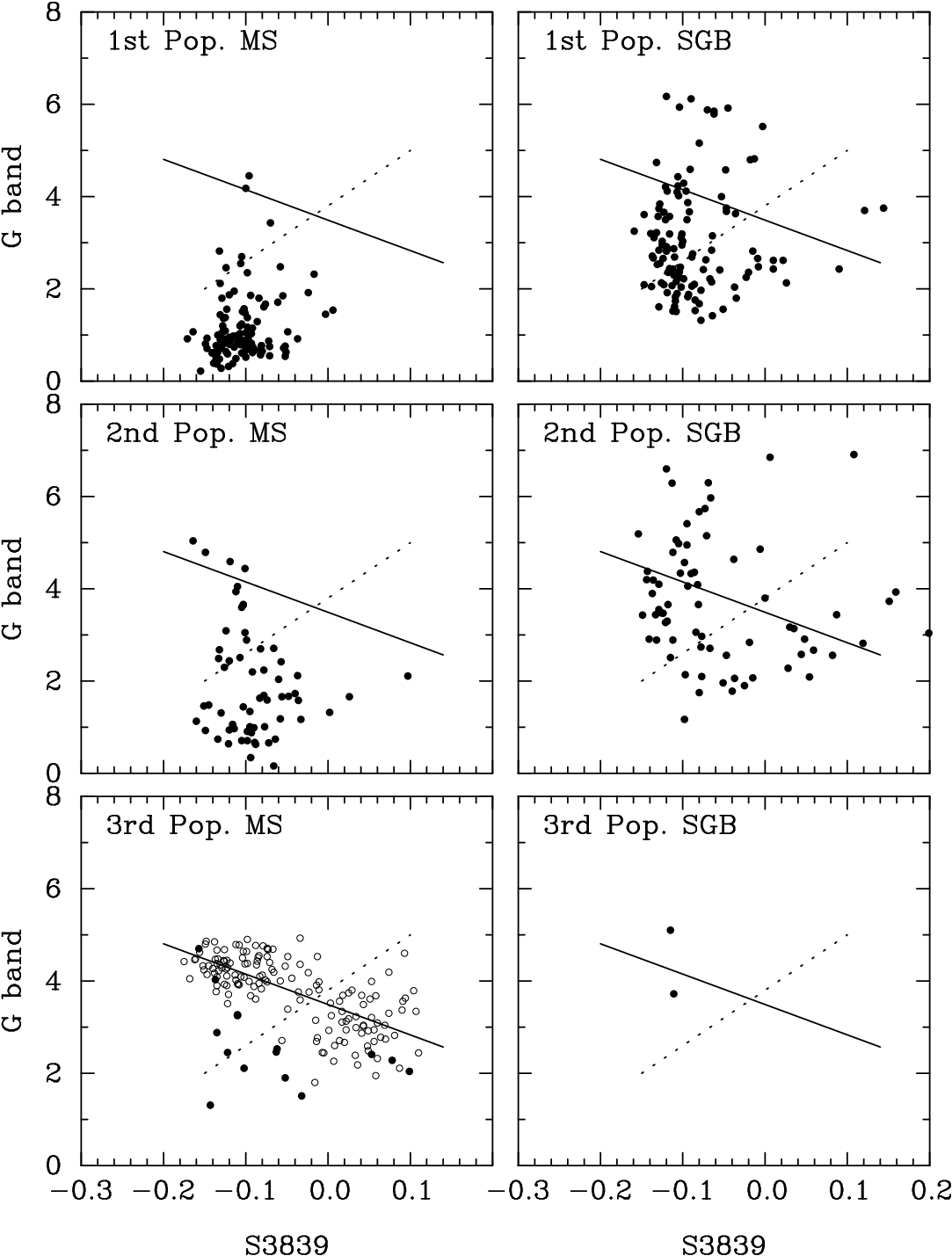} \figcaption[f5.ps]{CN vs CH
indices for {\wcen} (solid dots) and {\tuc} (open circles).  The
{\tuc} data, plotted only in the metal-rich, MS panel, are split
according to the CN/CH bimodality with the division shown by the
dotted line defined in the bottom left panel, and reproduced in the
other panels.  The solid line is a fit to the CN, CH anticorrelation
and plotted in each panel as a reference.  The {\wcen} data are
divided according to metallicity and position on the CMD (MS or SGB).
\label{cnch} }
\end{center}
\end{figure}

The CN/CH bimodality known to exist on the RGB of {\tuc} is also found
on the MS \citep{can98, dac04}.  CN strong and CN weak stars exist in
{\wcen} on the RGB (\citealt{nb75, per80, cb86}; \citetalias{nd95}),
but have not yet been seen on the MS in {\wcen}.  Figure \ref{cnch}
plots the CN index versus the CH one for the {\wcen} data, and is
split into the three metallicity bins as well as the MS and SGB groups
(left and right columns respectively).  Also plotted, as open circles,
are the {\tuc} data showing the bimodality within this cluster for
comparison with the metal-rich, MS population in {\wcen}. The solid
line is a fit to the {\tuc} anticorrelation and plotted in each panel
as a reference. The dotted line is the one used to define the CN
weak/CH strong, and vice versa, groups.

\section{First Pass Abundances} \label{abund_sect}

To obtain an initial estimate of the abundance ratios of stars in our
sample, in particular for the stars that have enhancements in carbon,
nitrogen and/or strontium, we compared indices obtained from synthetic
spectra with those from the observed data.  As described in
\S\ref{met_sect}, the {\wcen} sample was divided by metallicity into
three populations, as shown in Figure \ref{cmd2}.  Isochrones with the
mean metallicity and ages of each population were fitted to the data.
Discrete points along these isochrones gave a series of stellar
parameters (temperature and gravity) that are representative of the
observed data for each population.

Kurucz models (1993) were used with atomic line lists from R. A. Bell
(2000, private communication) and Kurucz molecular lines lists to
generate synthetic spectra to compare with our observations. The
spectrum synthesis code was originally developed by \citet{cn78}.  A
synthetic spectrum with appropriate stellar parameters was compared
with an atlas of the Sun \citep{beckers}.  Using an adopted solar C
abundance of log(N/N$_{tot}$)=--3.36, the CH feature showed a small
difference between the spectrum of the Sun, and the synthetic
spectrum, and as a result the log(gf) values were reduced by
0.5~dex. Similarly, for an adopted N abundance of
log(N/N$_{tot}$)=--4.04, the synthetic violet CN feature was too
strong with respect to the solar spectrum and the log(gf) values were
reduced by 0.25~dex. The log(gf) values for the blue CN feature
($\lambda$4215{\AA}) were reduced by 0.10~dex in a similar manner.  The
Sr {\sc ii} 4077 and 4215 lines and Ba {\sc ii} 4554 line were also
compared and found to be within acceptable agreement.  An observed
spectrum of Arcturus ($\alpha$ Boo) was then compared with a synthetic
spectrum with appropriate stellar parameters \citepalias{nd95} and was
used as an independent check of the adjusted values.  It was found to
provide acceptable agreement with abundance studies of this star.
Table \ref{tbl-2} lists the gf values and solar abundances used here
for Sr and Ba.

The derived abundance of C is influenced by the amount of oxygen
within the star as the CO molecule forms preferentially over the CH
one.  However, for our range of stellar parameters, changing the O
abundance was demonstrated to have no effect on our calculations.
This was tested by generating a series of synthetic spectra at
metallicities that spanned the {\wcen} metallicity range with several
temperatures and gravities typical of the observed sample. The oxygen
abundance was varied for each one (between [O/Fe]=0.0 and 0.5 dex),
and the strength of the G band was compared between synthetic spectra
with the same stellar parameters.  It was found that oxygen did not
play a significant role in the abundance determined from the CH
feature for any of the stellar parameters considered.  This test was
also conducted for the CN feature and again no noticeable difference
was found with varying oxygen abundance.  We have assumed that oxygen
follows the general $\alpha$-enhancement trend.  That is, [O/Fe]=0.3
for objects with metallicities [Fe/H]$\leq$--1.0, and [O/Fe]=+0.18 for
the higher metallicity stars.  \citetalias{nd95} show that there is a
number of stars with oxygen deficiencies (values of [O/Fe]$<$0.0) for
RGB stars.  This indicates that the ON cycle may have occurred in
these objects.  Although the assumed [O/Fe] abundance does not play a
significant role in the abundances determined, it would be beneficial
to determine the oxygen abundances for individual stars in
future studies.

The stellar parameters derived from the isochrones were used to
generate synthetic spectra for ten carbon abundances at each point
along an isochrone, for each population.  G band indices were
determined for each of these synthetic spectra and plotted along with
the indices from the observed spectra for each population.  This
enabled a mean carbon abundance to be obtained for each population,
which was then used when another set of synthetic spectra were
generated with the nitrogen abundance being the varying parameter.
The S3839 index was measured in these synthetic spectra and the
results for the synthetic spectra were again plotted with the observed
data to obtain a mean nitrogen abundance for each population.  This
procedure was followed again, but with varying strontium abundance and
the strontium indices plotted for the synthetic and observed data.

Interpolation between the synthetic isoindex lines for each element
and population enabled a first estimate at the abundances of carbon,
nitrogen and strontium for each object.  From these abundances the
objects with possible enhancements (or in some cases depletions) in
carbon, nitrogen and strontium could be identified for further in
depth analysis. The synthetic spectra give an initial estimate of the
abundances, but should be viewed with caution since there is a
0.4--0.6 dex range in {\feh} for the member stars of each population
but the isoindex lines are for a fixed abundance.  Also, some stars
have temperatures and gravities that deviate from those used for the
isoindex lines.  The larger the difference in {\feh}, temperature
and/or gravity from the isoindex value, the larger the uncertainty in
the initial estimate.

\begin{figure}
\begin{center}
\includegraphics[scale=.60,angle=0]{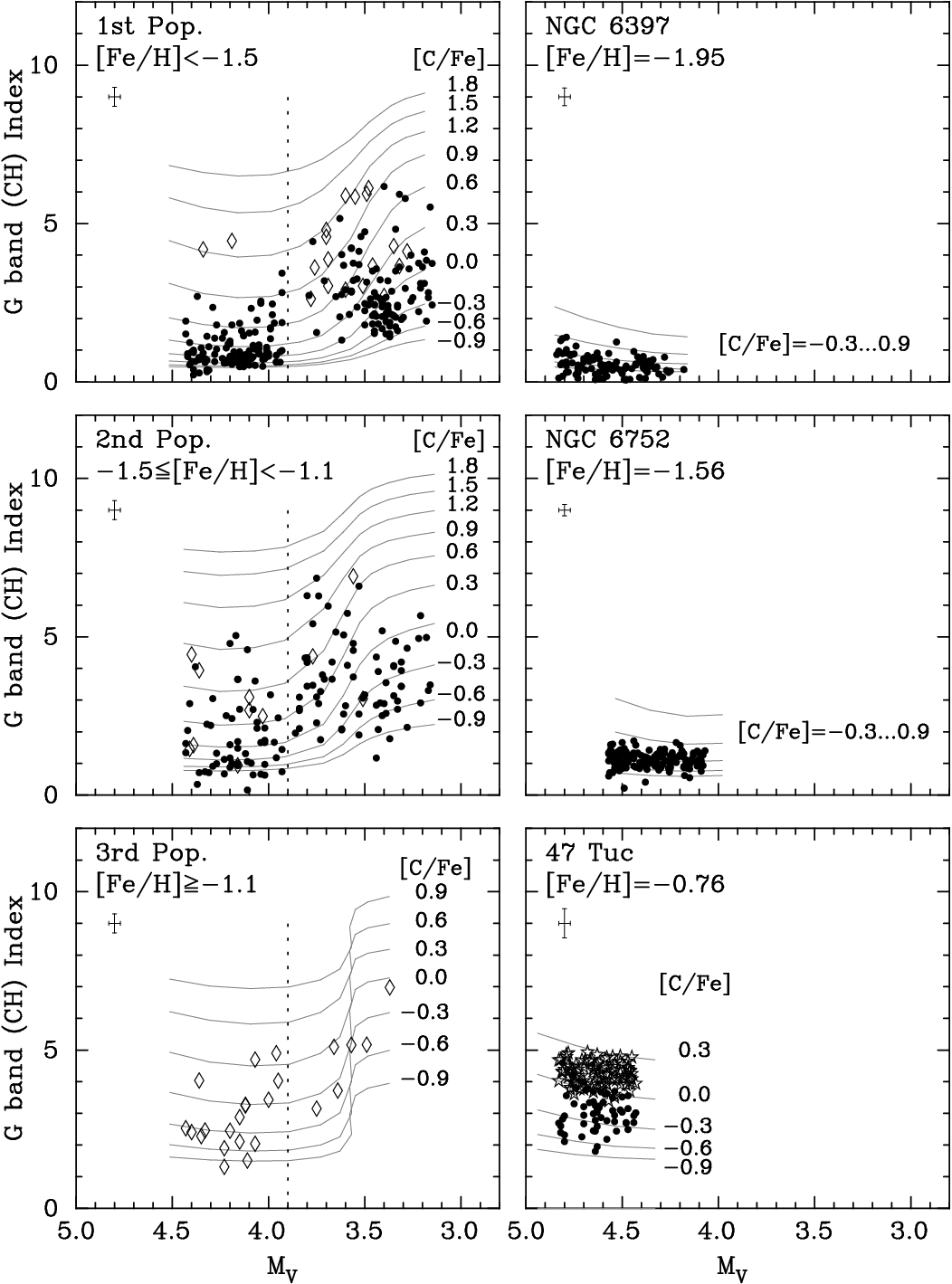} \figcaption[f6.ps]{G band
indices plotted against absolute magnitude (M$_{V}$) for each of the
three populations of {\wcen} in the left panels, and calibrating
globular clusters in the right panels.  The diamonds represent those
stars for which individual abundances were determined, while filled
circles represent all other objects.  Lines of common abundance are
also plotted for the synthetic spectra. The vertical dashed line
represents the cutoff between main sequence and turnoff stars. See
text for details.
\label{c1st} }
\end{center}
\end{figure}

The C abundance for the synthetic spectra was varied between
[C/Fe]=--0.9 and +1.8.  The G band index was measured for each
spectrum and plotted with the observed data, shown in the left column
of Figure \ref{c1st}.  The diamonds represent those stars for which
individual abundances were subsequently determined (see
\S\ref{iaa_sect}).  The synthetic data for each abundance are plotted
as solid lines.  A large spread in C abundance can be seen within both
the MS stars and the SGB stars and there are two MS stars that have
possible enhanced features.  The C abundance for the $\omega$ Cen
stars were calculated by interpolation in luminosity and G band
strength to give an initial approximation to the abundance of that
element.

The right column of Figure \ref{c1st} shows the three calibrating
globular clusters (NGC~6397, NGC~6752 and {\tuc}). NGC~6397 and
NGC~6752 bracket the metallicity range of the 1st Pop., and their
results as normal globular clusters can be compared to those found for
{\wcen} in order to determine which stars are {\it abnormal} on the MS
of the latter.  The error bars in each panel represent errors in
measured index and magnitude.  They do not include the uncertainty in
the inferred abundance induced by an object's distance in color,
magnitude and metallicity from the isochrone used to generate the
isoindex lines.

The normal globular clusters show no significant spread (1$\sigma$) in
the CH index, while the first population of {\wcen}, on the other
hand, shows several objects that are outliers from the bulk of the
population and a 2$\sigma$ spread on the MS and a 4$\sigma$ spread on
the SGB.  NGC~6752 can also be compared with the 2nd pop.  Again the
small spread in the normal GC is in contrast to that seen in {\wcen}.
The 3rd Pop. can be compared with the results from {\tuc}, for which
the indices have been split according to the CN--CH bimodality for
clarity.  On close inspection, one sees that the {\wcen} results have
a similar range in indices to {\tuc}.

\begin{figure}
\begin{center}
\includegraphics[scale=.60,angle=0]{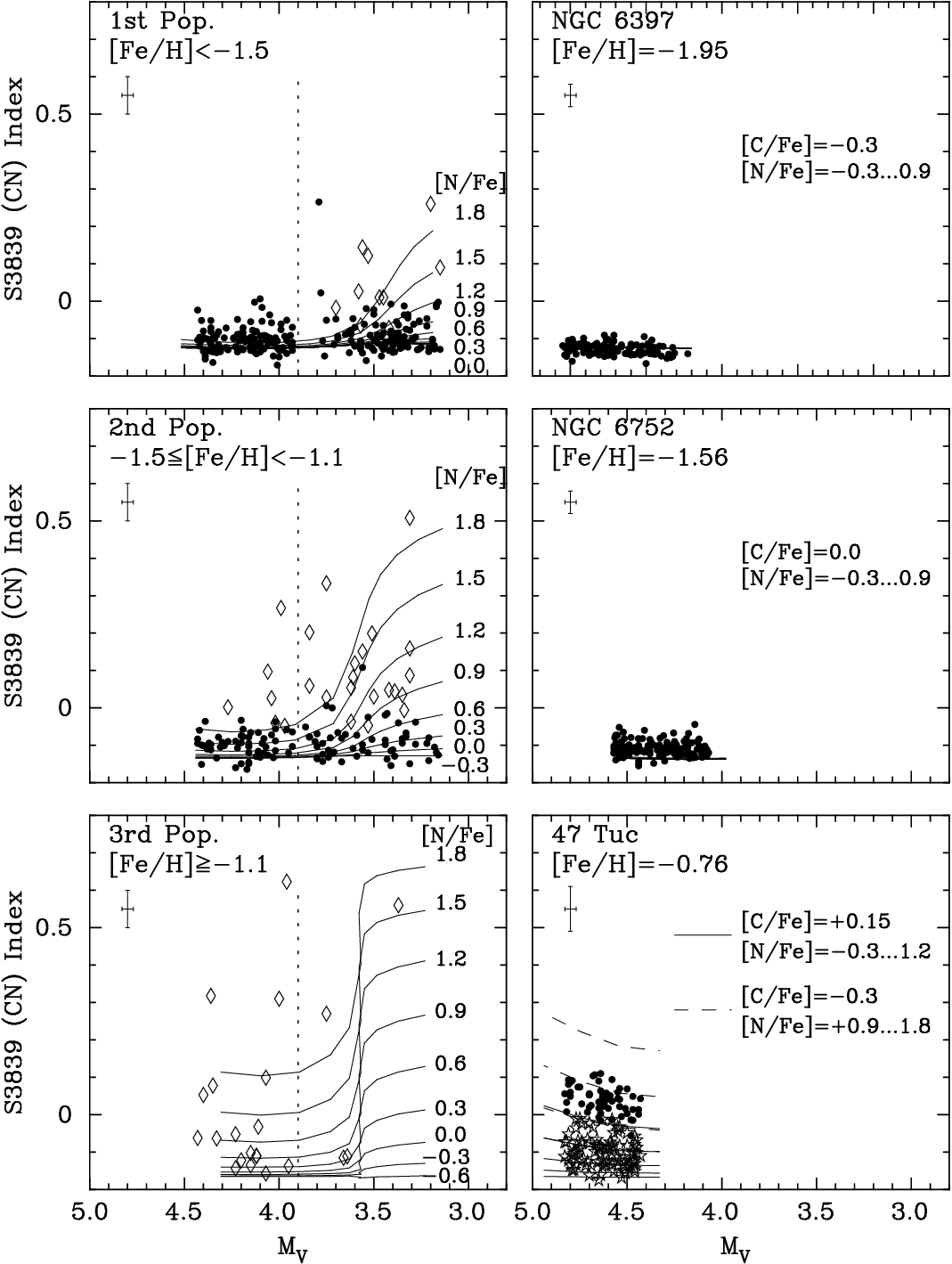} \figcaption[f7.ps]{CN
indices plotted against absolute magnitude (M$_{V}$) for each of the
three populations in the left hand panels and the calibrating globular
clusters in the right.  The diamonds represent those stars for which
individual abundances were determined, while filled
circles represent all other objects.  Lines of common abundance are
also plotted for the synthetic spectra. The vertical dashed line
represents the cutoff between main sequence and turnoff stars. See
text for details. \label{n1st} }
\end{center}
\end{figure}

The mean C abundance for the bulk of each population was determined
and then used when calculating synthetic spectra with varying
nitrogen.  Again, the CN index was determined for each synthetic
spectrum and plotted against the observed data, shown in the left
column of Figure \ref{n1st}. Once more, the error bars in each panel
represent errors in measured index and magnitude. The diamonds in this
plot indicate those stars that had individual N abundances determined
(see \S\ref{iaa_sect}).  As with the plot of the CH indices, the top
panel shows the most metal-poor stars, the middle one the intermediate
metallicity population, and the bottom panel the most metal-rich
stars.  Inspection of the top panel shows the synthetic lines for CN
are closely spaced on the MS, showing the low sensitivity of the CN
feature at these abundances and temperatures. Consequently, finding
any objects with high N abundances is difficult, and any spread is
comparable to the measurement errors (1$\sigma$).  Determining N
abundances becomes a slightly easier task on the SGB and the presence
of several outliers is clear. The outliers are more pronounced in the
2nd Pop. and there is a large range in apparent N abundance in the 3rd
population with 4$\sigma$ and 9$\sigma$ spreads on the MS and SGB,
respectively.  Initial estimates of the N abundance for each {\wcen}
star were calculated, and all objects with a first pass N abundance,
[N/Fe], greater than 1.2~dex were included for further study, although
this includes a fair number of objects in the metal-poor, MS category
that were subsequently shown to have normal N abundances due to the
low sensitivity of the feature.

The right column of Figure \ref{n1st} shows the results for the
calibrating globular clusters.  Both NGC~6397 and NGC~6752 show little
scatter in their indices, as would be expected when the synthetic
indices are taken into account.  There is very little sensitivity in
this feature at these metallicities, temperatures and gravities.
{\wcen}, as with the G band, shows several stars that have CN indices
that are higher than the mean, even with the low sensitivity in this
area.  The indices for {\tuc} are again split into the CN--CH
bimodality.  Two sets of synthetic lines were plotted for these data,
with different mean C abundance. We see a larger range within the
{\wcen} data than for the {\tuc} indices.

\begin{figure}
\begin{center}
\includegraphics[scale=.60,angle=0]{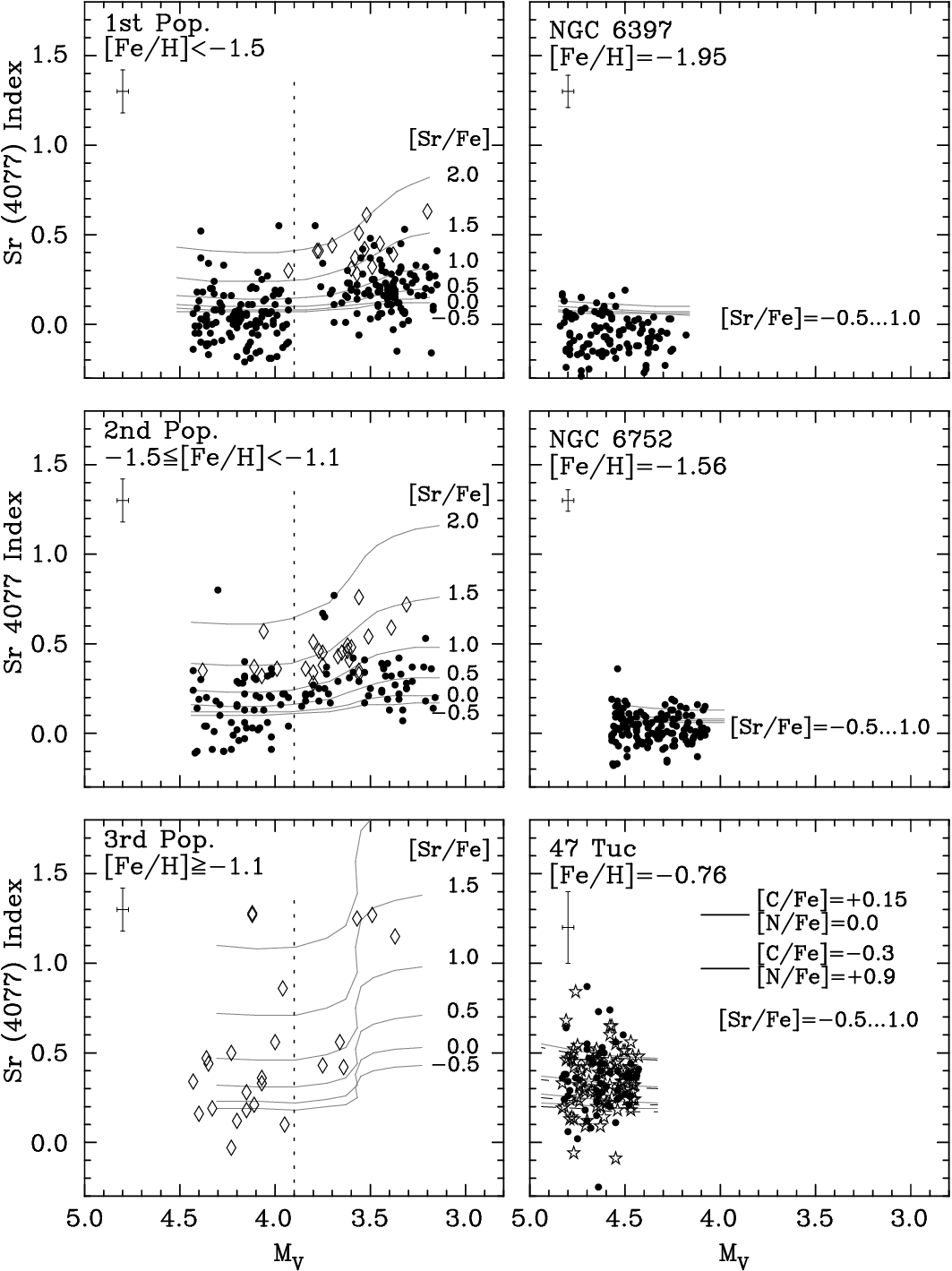} \figcaption[f8.ps]{ Sr {\sc
ii} 4077{\AA}, indices plotted against absolute magnitude (M$_{V}$)
for each of the three populations in the left hand panels and the
calibrating globular clusters in the right.  The diamonds represent
those stars for which individual abundances were determined, while
filled circles represent all other objects. Lines of common abundance
are also plotted for the synthetic spectra.  The vertical dashed line
represents the cutoff between main sequence and turnoff stars. See
text for details.\label{s1st}}
\end{center}
\end{figure}

Sr and Ba were analyzed simultaneously, although the Ba index was too
weak to obtain meaningful results. The synthetic Sr indices are
plotted in the left column of Figure \ref{s1st} along with the {\wcen}
data, where the error bars in each panel represent errors in measured
index and magnitude.  A large spread of the Sr index is apparent in
Figure \ref{s1st}, and is mainly due to the low S/N and intermediate
resolution of our spectra.  The spread in the measured index is
comparable to the measurement error in the first two populations.  The
third population shows a 3$\sigma$ spread in comparison with the
measured errors.  Despite this, we are able to identify objects that
may have large Sr enhancements, and use these for individual spectrum
synthesis analysis.  The diamonds represent those objects that had
individual Sr abundances calculated, which are discussed in
\S\ref{iaa_sect}.

The normal globular clusters all show no Sr enhancement and any
scatter is comparable (of order 1$\sigma$) to the measurement errors
of the index.  {\tuc} shows several objects that may be considered to
have higher Sr abundance ratios, but closer inspection reveals that
these objects have higher indices due to low S/N.  {\wcen}, in all
cases, shows larger scatter than the normal globular clusters, with
several stars having higher indices being objects of particular
interest.  Inspection of spectra shows these are not the result of
lower S/N.

The choice of isochrones that were used for each of the three
populations in the first pass abundance analysis was determined from
the mean metallicity and ages of the respective populations. The
impact of a different choice of parameters for the three sets of
isochrones was evaluated. The first population was assigned an
isochrone with [Fe/H]=--1.9 and age=15 Gyr (as against --1.7 and 13.5
Gyr), the second population an isochrone with [Fe/H]=--1.2, age=11 Gyr
*cf. --1.4 and 13.0 Gyr) and the third an isochrone with [Fe/H]=--0.6
and age=9 Gyr (cf. --0.9 and 12 Gyr).  Most stars showed differences
in first pass carbon abundances of less than 0.5~dex.  Strontium
abundances were calculated with differences less than 0.8~dex.
Differences in nitrogen values were affected the greatest of the three
elements, with differences ranging up to $\pm$2.0~dex in the most
extreme case, and $\pm$1.0 on average.  In practice, the result for
the nitrogen abundances was not significant as objects with enhanced N
selected for further analysis were determined using the S3839 index
rather than the derived N abundance, as described in the following
section.  The use of a different distance modulus and reddening were
also investigated and the differences in derived abundance was found
to be less than 0.2~dex for carbon and nitrogen, and less than 0.5~dex
for strontium.  These differences are all within the error margins
placed on the selection of stars with possible enhancements in C, N or
Sr.

\section{Individual Abundance Analysis} \label{iaa_sect}

\subsection{Abundance Determination of [C/Fe], [N/Fe] and [Sr/Fe]}

Following the analysis in \S\ref{abund_sect}, the subsample of stars
with possibly enhanced C, N or Sr abundances were individually
analyzed using spectrum synthesis techniques.  These objects were
chosen according to their first pass abundances.  For carbon, if a
star's interpolated abundance was [C/Fe]$>$0.3 it was included.  The
cutoff for Sr was [Sr/Fe]$>$0.8.  The CN index was more complicated
due to the weakness of the feature and hence its low sensitivity to C
and N.  Any interpolated abundance had very large error bars due to
the small separation between isoindex lines.  Therefore, these objects
were chosen based on the S3839 index rather than interpolated
abundance and the cutoff imposed was S3839$>$--0.05.  All objects in
the 3rd Pop. were included for individual analysis.

Stars with [C/Fe]$<$--0.75 from the first pass abundance estimation
were also included from the first two populations. However it was
found from the comparisons of observed and synthetic spectra for
individual stars that in most instances there was not sufficient
sensitivity in the data to reliably measure depletions.  The low
sensitivity of the CN and Sr features on the MS also meant that it was
not possible to measure such depletions in N and Sr, respectively, and
as such we can not say anything about whether depletions exist amongst
these populations.

An age was assigned to each star based on its position on the CMD and
metallicity using theoretical isochrones \citep{yi01} as described in
\citet{sta06a}.  It was relatively straightforward to also obtain
temperatures and gravities from the isochrones simultaneously with the
ages. A grid of isochrones was used which span the metallicity range
--2.6$<$[Fe/H]$<$0.3 in 0.05 dex increments. For each metallicity
there were 34 isochrones with ages 2--19~Gyrs in 0.5~Gyr steps. Alpha
enhancement was taken to be constant ([$\alpha$/Fe]=0.3) for
[Fe/H]$\leq$--1.0, and declining linearly for higher [Fe/H] until it
reached the solar value at [Fe/H]=0.  To assign an age to each star,
its metallicity was used to select the nearest isochrone in our
grid. The isochrones with this metallicity but with differing ages
were then compared to the star's $(B-V)_0$ and M$_{V}$ on the CMD to
find the one closest. Usually a star's position did not fall directly
on one isochrone and linear interpolation in color or magnitude was
performed between the two closest ones to determine its age, along
with the temperature and gravity.  In almost every case the derived
temperature and gravity are not the same as those from the average
isochrones used in Figure \ref{cmd2}. Comparisons were made with
temperatures obtained from color temperature relations in
\citet{alo96}. These were, on average, smaller by 100K.  This value is
within the 150K uncertainty in temperature used in the error analysis.
Microturbulence was assumed to be 1.0 kms$^{-1}$.

\begin{figure}
\begin{center}
\includegraphics[scale=.60,angle=0]{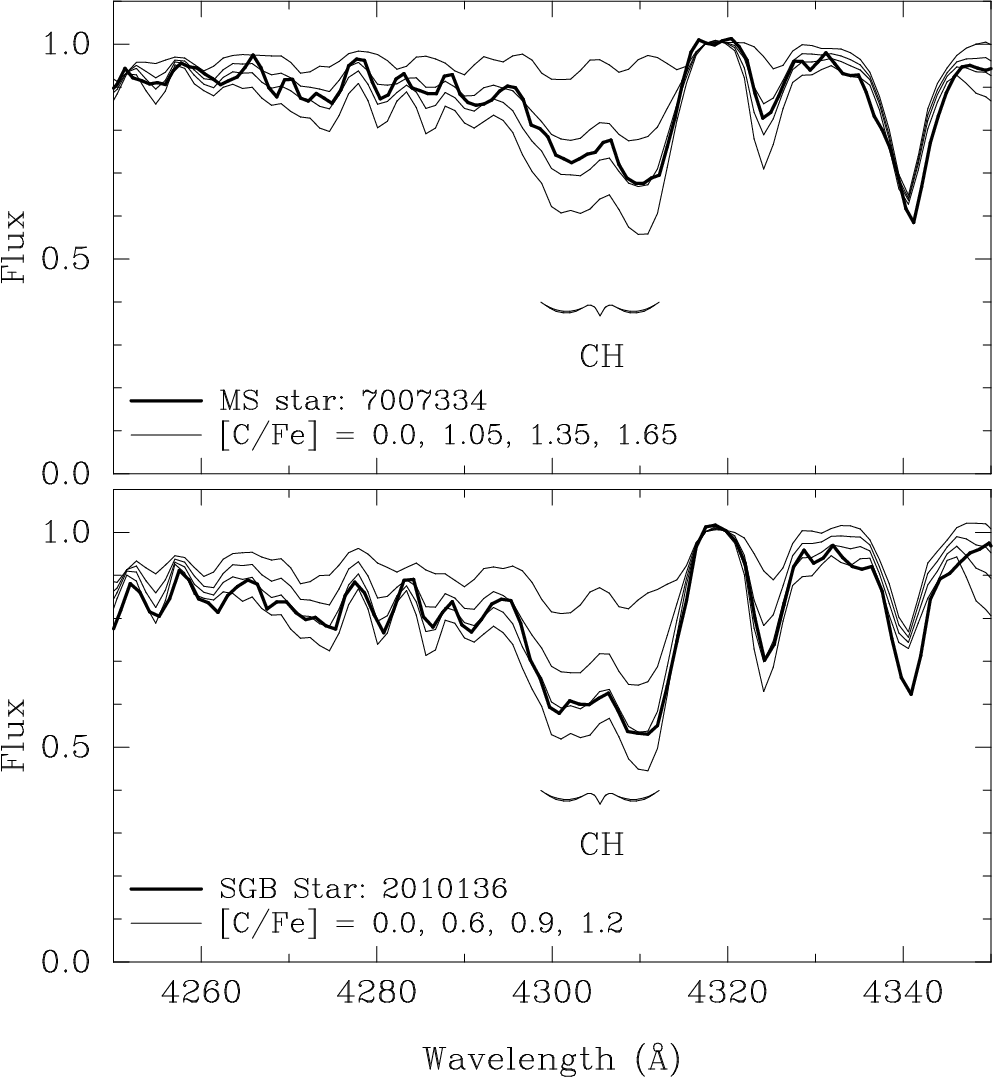}
\figcaption[f9.ps]{Examples of observed and synthetic spectra for two
objects with enhanced carbon abundances.  The bold line represents the
observed spectrum in each case.  The top panel shows an example of a
MS star, 7007334, with [C/Fe]=1.35, while the lower panel shows a SGB
star, 2010136, with [C/Fe]=0.9.  Synthetic spectra with four carbon
abundances are shown for each object, including [C/Fe]=0.0 for
completeness.
\label{cspec} }
\end{center}
\end{figure}

Each candidate had the abundances of C, N and Sr determined.  The C
feature (CH) at $\sim$4300{\AA} was analyzed first.  Figure
\ref{cspec} gives examples of a MS star (upper panel) and SGB star
(lower panel) with enhancements in carbon.  The heavy, solid lines
depict the observed spectrum, while the light solid lines show the
synthetic spectra.  The carbon abundance ratio was modified in the
synthetic spectra while all other parameters were held constant.  The
determined carbon abundance ratio is plotted, along with two others
bracketed the value by $\pm$0.3~dex.  A synthetic spectra having
[C/Fe]=0.0 is also shown for comparison.

\begin{figure}
\begin{center}
\includegraphics[scale=.60,angle=0]{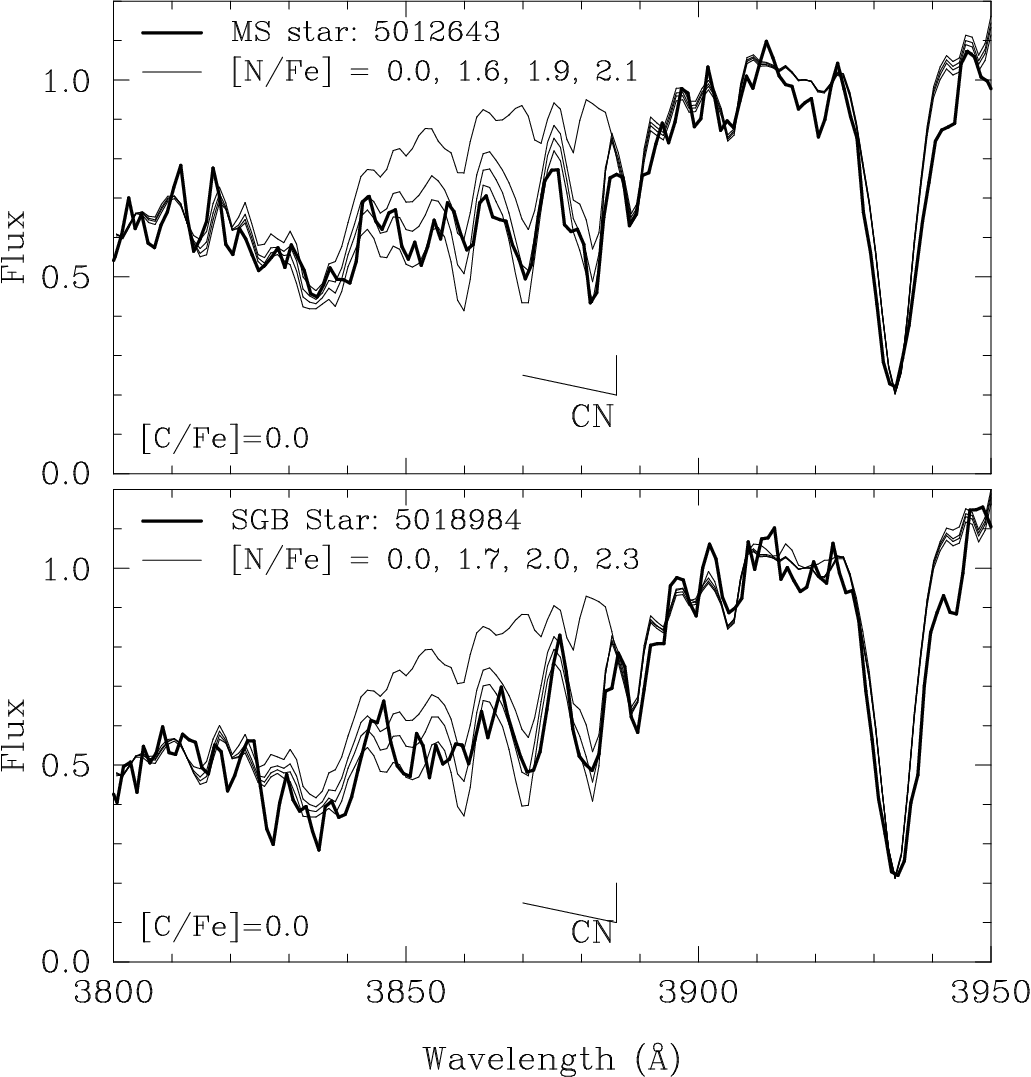}
\figcaption[f10.ps]{Example observed and synthetic spectra of two
objects with enhanced nitrogen abundances.  The top panel shows the
spectrum of a MS star in bold, 5012643, and synthetic spectra as
light lines for a range of nitrogen abundances including [N/Fe]=0.0.
The bottom panel shows a SGB star, 5018984, with [N/Fe]=2.0~dex. For
both stars, [C/Fe]=0.0.
 \label{nspec} }
\end{center}
\end{figure}

Given [C/Fe], the CN feature at $\sim$3883{\AA} was used to obtain the
N abundances.  It proved to be quite difficult to obtain accurate
abundances at the resolution of our spectra.  There was little
sensitivity of the feature at abundances lower than [N/Fe]=1.0~dex,
especially for objects with low {\feh}. Therefore abundances were only
recorded when [N/Fe]$\geq$1.0~dex.  Figure \ref{nspec} gives examples
of a MS star (top panel) and SGB star (lower panel) with enhancements
in nitrogen.  The heavy, solid lines indicate the observed spectrum,
while the light solid lines show the synthetic spectra.  The
determined nitrogen abundance ratio is plotted, along with two others
bracketing the value by $\pm$0.3~dex.  A synthetic spectrum with
[N/Fe]=0.0 is also shown for comparison.

\begin{figure}
\begin{center}
\includegraphics[scale=.70,angle=0]{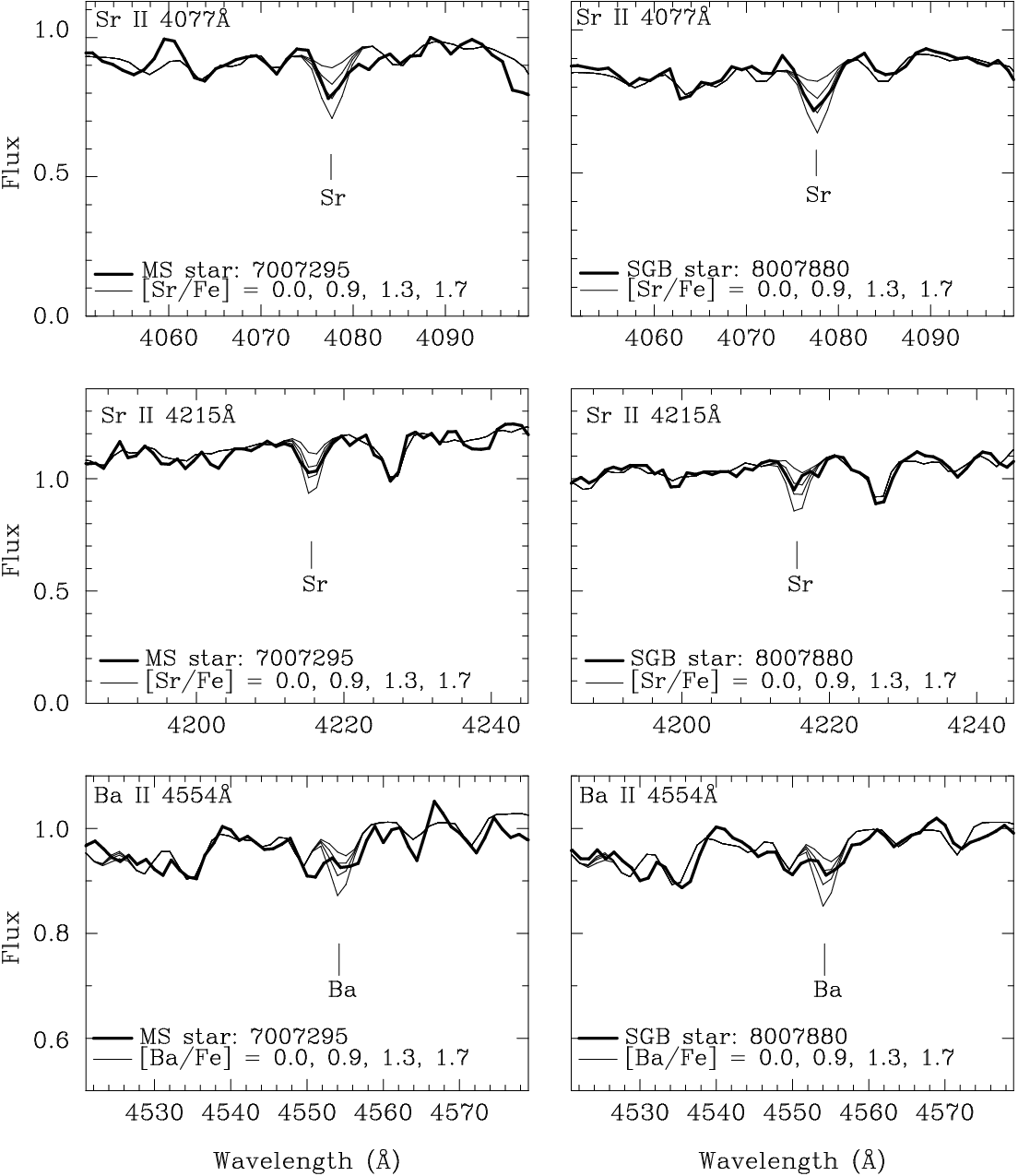}
\figcaption[f11.ps]{Observed and synthetic spectra showing a MS
object, 7007295, on the left. The right panel shows a SGB star,
8007880, with enhanced s-process elements.  The top panel in each
shows the Sr {\sc ii} 4077{\AA} line, the middle shows Sr {\sc ii}
4215{\AA} and the bottom shows \mbox{Ba {\sc ii} 4554{\AA}}.
\label{sspec} }
\end{center}
\end{figure}

The Sr abundance was obtained from the Sr {\sc ii} line at 4077{\AA}
line.  However, at the S/N of the data it was indistinguishable from
the noise at low abundances and results were only recorded when
[Sr/Fe]$\geq$1.0~dex.  Further, the abundance obtained from this line
was only accepted if the (weaker) Sr {\sc ii} line at 4215{\AA}
confirmed the abundance of the first line.  The Ba {\sc ii} at
4554{\AA} line was also examined. However its sensitivity was too low
to obtain reliable abundances in most cases.  It was able to be used
as an extra confirmation of the Sr abundance, as the Ba abundance is
expected to be strongly correlated with that of Sr at these
metallicities.  Figure \ref{sspec} gives examples of a MS star (left
column) and SGB star (right column) with enhancements in strontium.
The heavy, solid lines indicate the observed spectrum, while the light
solid lines show the synthetic spectra. The top panels show the Sr
\textsc{ii} 4077{\AA} line, the middle panels the Sr \textsc{ii} 4215
{\AA} one.  The strontium abundance ratio determined for these stars
is plotted, along with two others bracketing the value by
$\pm$0.3~dex.  A synthetic spectrum with [Sr/Fe]=0.0 is also shown for
comparison.  The lower panels show the Ba \textsc{ii} 4554{\AA} line
with the same enhancements in [Ba/Fe] as used for [Sr/Fe].

\subsubsection{Errors}

Typical errors were calculated for several objects with different
metallicities.  The temperature, gravity and metallicity were varied
independently.  The measured index was also varied by its errors and
included in the error calculations.  The temperature was varied by
150K, determined by reassigning an isochrone to an individual star
based on the error in photometry.  The gravity varied by 0.2~dex
(determined in a similar way to the uncertainty in temperature), and
the metallicity by the individual abundance error for the star being
analyzed.  The error contributions were added in quadrature.  Typical
errors were 0.27~dex for the C abundance, 0.37~dex for the N abundance
and 0.38~dex for Sr.  Note that these errors refer only to the objects
that had abundances determined for them.

Abundances were also calculated for a small number of {\it normal}
stars across the three metallicity populations to verify that we would
obtain unenhanced abundances for these objects.  This was found to be
the case, but again the limited sensitivity of the CN and Sr features
may not reveal possible enhancements of N and Sr, respectively.

\subsubsection{Results}

Our abundances are presented in Table \ref{tbl-3}, which gives the
identification number for each star in column 1, and the V magnitude
and B--V in columns 2 and 3.  The temperature, gravity and metallicity
are listed in columns 4--6. In columns 7--10 indices for the G band
(CH), S3839 (CN), Sr 4077{\AA} and Ba 4554{\AA} are listed.  Abundance
ratios of C, N and Sr, relative to iron, are given in columns 11, 12
and 13, respectively, if calculated.  To untangle the abundance
patterns, each metallicity population was treated separately and
abundance information then compared between populations.  The
positions, photometry, metallicity and ages for these stars are
available in the electronic version of Table 2 in \citet{sta06a}.

When performing the spectrum synthesis analysis, comparisons by eye
between the Ca \textsc{ii} K line in the observed and synthetic
spectra were used as a check on the metallicities and temperatures.
In a small number of cases there was a large discrepancy between the
observed and synthetic, and therefore those stars were not considered
further.  In total, 32 stars did not have an abundance analysis
performed due to unreliable metallicities and/or ages.  It is unclear
whether the source of the error is driven by the metallicities,
related to a 3$\sigma$ error in the photometry or both.  After
excluding these stars, this left a total of 392 stars in our sample.

Table \ref{tbl-4} gives the number and percentage of normal, C, N and
Sr enhancements for each metallicity population.  Due to the varying
sensitivity of the CH, CN and Sr features, the emphasis of this paper
is on objects with enhancements: [C/Fe]$\geq$0.5~dex,
[N/Fe]$\geq$1.0~dex, and [Sr/Fe]$\geq$1.0~dex, even though abundance
ratios were calculated $\sim$1$\sigma$ below these values. These
limits were put in place to ensure we had complete numbers of objects
within our sample above these values.  There are objects with
enhancements in carbon and/or nitrogen, and strontium which therefore
are counted in both the carbon/nitrogen enhanced and strontium
enhanced groups.  This means the percentages often sum to more than
100\%.

Using a Gaussian distribution with a kernel equal to the error in
each abundance, the effect of errors on the number of objects found to
be enhanced was investigated.  The first distribution was centered at
[C/Fe]=0.0, with $\sigma_C$=0.27~dex, and produced 5\% of the sample
as enhanced objects under our criteria of [C/Fe]$\geq$0.5.  This is
comparable to the percentage of enhanced objects found on the
MS. However, the two metal-poor objects on the MS (9005309 and
7007334, with [C/Fe]=1.30 and 1.35, respectively) have $\sim5\sigma$ C
enhancements and are not likely to result from the uncertainties.  The
number of objects on the SGB that were found to be enhanced is
significantly higher than 5\%, but may be due to the bias against the
bulk of the metal-poor objects in the SGB sample introduced by the
blue cutoff at (B--V)=0.6 for the 2002 sample.

A similar procedure was employed for the nitrogen and strontium
abundances, using $\sigma_{N}$=0.37~dex and
$\sigma_{Sr}$=0.38~dex. Both the N and Sr simulations indicated that
it would be unlikely that any objects would be included in the
enhanced sample as spurious detections, since less than 1\% of the
simulated sample had abundances greater than the criteria of
[N/Fe]$\geq$1.0~dex and [Sr/Fe]$\geq$1.0~dex.

Table \ref{tbl-5} lists the abundance patterns found on the main
sequence and subgiant branches.  One should note that often when we do
not find any stars of a particular abundance pattern, it is due to low
sensitivity of the relevant observed features. The table only lists
numbers of stars with detectable enhancements of [C/Fe]$\geq$0.5~dex,
[N/Fe]$\geq$1.0~dex and/or [Sr/Fe]$\geq$1.0~dex.  Due to the selection
of the regions on the CMD, meaningful comparisons between the MS and
SGB are not possible.  The blue color limit for the SGB box is redder
than the bulk of the metal-poor population in the cluster.  Due to
these selection effects there are probably more detections of unusual
stars on the SGB than on the MS.

\subsection{Abundance Patterns}

Figure \ref{abpat} plots the abundance ratios (relative to iron) of
carbon, nitrogen and strontium for each of the three populations, as a
function of luminosity.  Care must be taken with interpretation of
this diagram. The shaded regions in the figure show the areas of
limited sensitivity and are plotted below the cutoff values by an
amount of $\sigma$/2.  Consequently, we do not detect all the
enhancements/depletions that are potentially present in our sample.
Therefore, only a few conclusions can be drawn from the recorded
abundance ratios.  Stars have been plotted with different symbols in
order to better facilitate the comparisons between different elements
and abundances.  For the 1st and 2nd populations the open triangles
represent those stars that are enriched in carbon ([C/Fe]$\geq$0.5)
and strontium ([Sr/Fe]$\geq$1.0).  Objects that are enhanced in both
nitrogen and strontium are shown by solid stars.  The one object with
enhancements in carbon, nitrogen and strontium, 8001811, is
represented by a plus sign inside a circle.  All other stars are
represented by solid dots with no cross-referencing between panels.
Since all stars in the 3rd population have abundance determinations,
the symbols are used to show the corresponding object between carbon
of any value with enhanced Sr and unenhanced N (as open triangles), or
enhanced N and enhanced Sr (as solid stars).  Solid triangles
represent stars with carbon abundances and nitrogen enhancements that
do not have high Sr abundances.

\subsubsection{Carbon} 

\begin{figure}
\begin{center}
\includegraphics[scale=.70,angle=0]{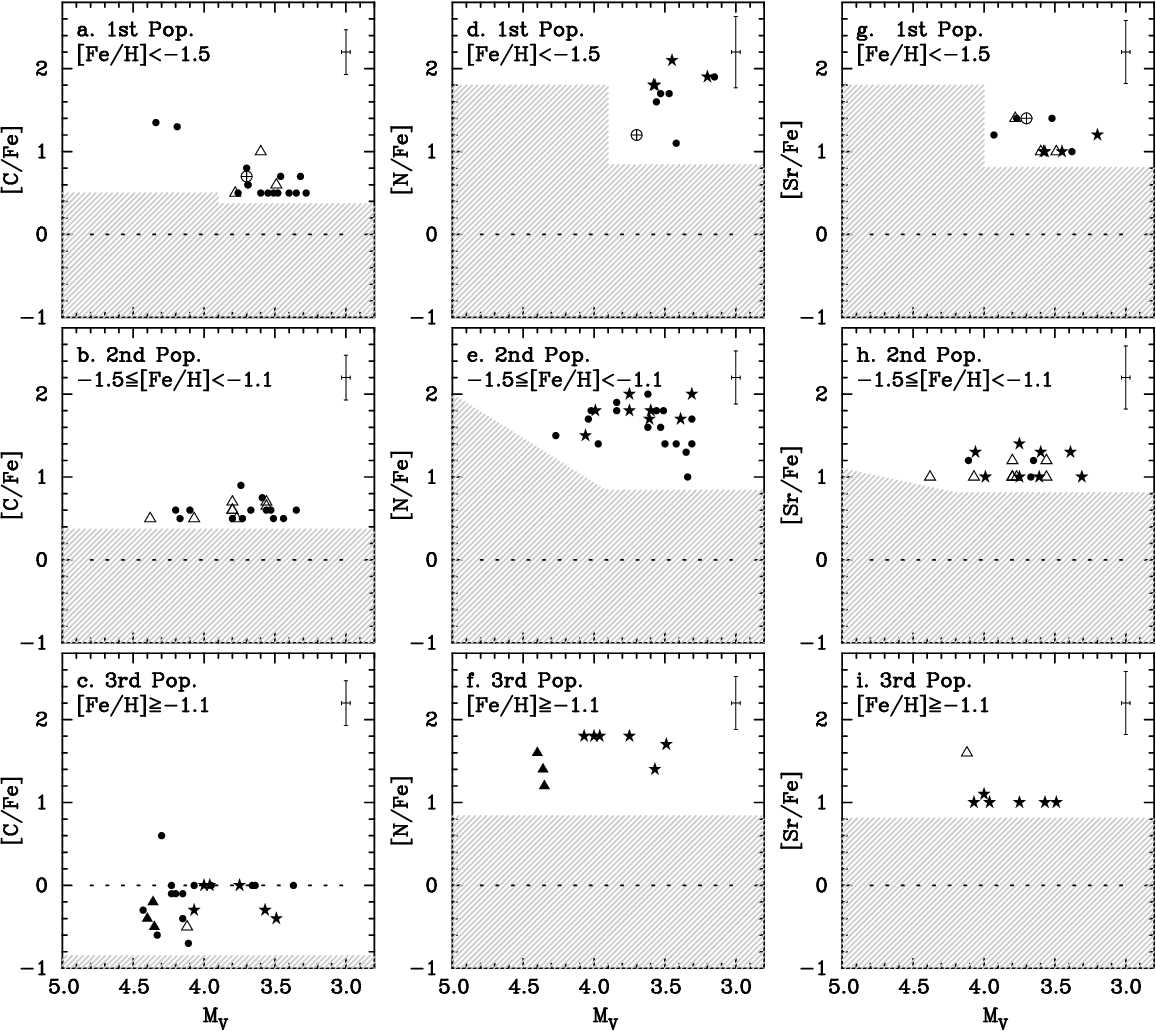} \figcaption[f12.ps]{
Abundance ratios (relative to Fe) of C, N and Sr for the three
metallicity populations as a function of magnitude.  The shaded
regions indicate areas of low sensitivity where the abundances are not
able to be determined reliably, and as such care should be taken in
interpretation of the abundance patterns. The dotted horizontal line
represents solar abundance ratios. For main sequence stars V$>$18
(M$_{V}>$4.04) and turnoff stars V$\leq$18 (M$_{V}\leq$4.04).  Also,
one should note the limited sensitivity of the CN and Sr features,
particularly at low metallicities, which makes it difficult to
recognize all objects of interest.  Finally, the biased color
selection of the SGB sample that excludes many metal-poor stars should
not be overlooked.  In particular, comparisons between the MS and SGB
groups should not be made.  See text for more details for definitions
of the symbols.
\label{abpat} }
\end{center}
\end{figure}

Panels a, b and c of Figure \ref{abpat} show the [C/Fe] abundances for
each population.  The objects in Fig. \ref{abpat} with individual
abundance determinations are shown as diamonds in the carbon, nitrogen
and strontium indices plots (Figures \ref{c1st}, \ref{n1st} and
\ref{s1st}). The reader is reminded that the spectrum synthesis was
performed for the entire 3rd Pop. sample, but only for a select group
of {\it enhanced} objects for the 1st and 2nd Pops.  Inspection of the
C enhanced objects in the 1st Pop. (Fig. \ref{abpat}, panel a) shows
there are relatively more objects with elevated abundances on the SGB
than on the MS.  This same effect is seen to a lesser degree in the
2nd Pop. data in Fig. \ref{abpat}, panel b.  If the process that led
to the formation of the C-rich stars was primordial (i.e.  they formed
from matter already enhanced in C and therefore have uniform
abundances throughout) one would expect equal numbers of these objects
on the MS and SGB.  On the other hand, if these objects had elevated C
abundances due to the accretion of material onto their surface layers,
as they move off the MS and onto the SGB the convective layer deepens
and such enhancements become diminished. Consequently, one would
expect not to find as many SGB objects with C enhancements,
contradicting the results found here.  Instead, the result may be due
to the diminished sensitivity of the CH feature on the MS compared
with the SGB preventing detection of enhanced C in main sequence
objects.

The metal-poor objects on the main sequence with large carbon
enhancements are possible counterparts of the CH stars found on the
RGB in {\wcen}.  Two objects on the main sequence, 7007334
([C/Fe]=1.35) and 9005309 ([C/Fe]=1.30), seen in Fig. \ref{abpat}
(panel a), show the largest carbon enhancement in our sample, with no
other metal-poor main sequence objects showing any enhancement. These
objects also show no detectable enhancements of nitrogen or strontium.
The study by \citet{bd74} found that for two of the CH giants in
{\wcen} (ROA 55 and 70), [C/H]$\sim$--0.8 and [N/H]$\sim$0.0. Adopting
[Fe/H]=--1.9 for ROA 70 \citep{gra82}, this gives [C/Fe]$\sim$1.1 and
[N/Fe]$\sim$1.9.  The carbon enhancements are similar to the values
found here, although the N enhancement is apparently not, though are N
sensitivity at this [Fe/H] is low.  As for the s-process elements,
\citet{gra82} found [s/Fe]=1.1 for ROA 70 (where s represents the
averaged abundances of the s-process elements), but for our MS stars
we do not detect any enhancement of Sr above [Sr/Fe]=1.0~dex.  A
metal-poor SGB star, 8001811, shows enhancements in carbon, nitrogen
and strontium, similar to those found for RGB CH stars in {\wcen}.
This star may also be similar to the ``subgiant CH'' stars reported by
\citet{lb91} in the field, that exhibit carbon and s-process
enhancement.

Figure \ref{c1st} indicates that there are some stars in each of the
three populations with possible C depletions.  Upon further
investigation it was found that in the first population with the
lowest metallicities, C abundances were not able to be reliably
determined below [C/Fe]=0.0 for the majority of candidates.  For the
second metallicity population, with \mbox{--1.5$\leq${\feh}$<$--1.1},
[C/Fe] started to lose its sensitivity around --0.2.  C abundances for
the most metal-rich population, with {\feh}$\geq$--1.1, were able to
be determined down to [C/Fe]$\sim$--0.7.

It is significant that we only find one object in the 3rd Pop. with C
enhancements greater than 0.0~dex.  All objects within the 3rd
Pop. went through the individual abundance determination process,
although again the CN and Sr features of some objects were not useful
for abundance determination. Most of the objects in the 3rd Pop.  were
C depleted, as can be seen in the panel c of Fig. \ref{abpat}.  Some
of these C depleted stars have various large N and/or Sr enhancements.
However, it is interesting to note that we find several objects (4/9)
with C depletions that do not have corresponding N and/or Sr
enhancements. This may be due to the low sensitivity of CN and Sr
features. 

\subsubsection{Nitrogen}

Inspection of the middle column of Figure \ref{abpat} shows the [N/Fe]
abundance ratios as functions of luminosity for each of the three
populations.  The limited sensitivity of the CN feature is indicated
by the shaded regions, and for this reason only the extremely N
enhanced objects are identified.  On the SGB in the metal-poor and
metal-intermediate populations it can be seen there is a fraction
($\sim$20\%) of N enhanced objects, with [N/Fe] up to $\sim$2.0~dex.
Similarly, the 3rd Pop. contains a significant fraction of objects
with N enhancements, of order 50\%.  These N abundances are
$\sim$0.5--1.0~dex higher than those found for RGB stars in
\citetalias{nd95}, and up to $\sim$0.75~dex higher than in the
analysis of RGB stars of \citet{bw93}.  This large discrepancy
between the present work and other studies may be due to differing
analysis techniques.  Analysis of RGB stars using the method employed
here would be useful in resolving this discrepancy.

Nitrogen enhancements are not seen in the 1st Pop. on the MS, although
such stars are seen on the SGB at this metallicity and on the MS for
the higher abundance groups. Whether this is due to the low
sensitivity of the CN feature at the resolution and S/N used here and
our subsequent inability to detect the objects with overabundances in
N, or is a real effect remains unclear.  That said, canonical stellar
evolution results do not predict the processing of C to N between the
MS and SGB phases of such low mass stars. Only one star on the SGB was
found with [N/Fe]$\geq$2.0.  Therefore the fact that no enhancements of
[N/Fe]$>$2.0 were seen for MS stars may be a statistical effect.
Higher resolution data are needed to address this question.
\citet{pio05}, in a study of the double main sequence, found nitrogen
enhancements of [N/Fe]=1.0 for the metal-intermediate bMS and
[N/Fe]$<$1.0 for the metal-poor rMS.  Therefore it seems likely that
our technique, which could only measure enhancements above [N/Fe]=1.0,
did not have the required sensitivity to measure any enhancements in N
on the main sequence.

\subsubsection{Strontium}

For Sr there is a abundance trend similar to that found for N across
the three populations, as may be seen in Figure \ref{abpat} (right
column).  Here one finds a significant number of objects that show Sr
enhancement in all three metallicity populations.  In particular, we
draw attention to the most Sr-rich object, the main sequence star
2015448, which has [Sr/Fe]=1.6.  This object has the added anomalous
behavior that while the abundance of Ba usually tracks that of Sr in
heavy neutron-capture enhanced objects, we do not detect Ba, and
estimate [Ba/Fe]$<$0.6.  This object is discussed in more detail in
\citet{sta06b}.

\subsubsection{Metal rich Population}

The 3rd Pop. differs from the other two populations in that there were
essentially no enhancements in C across the MS or SGB. One 3rd
Pop. object, 2016543, was identified with mild C enhancement.  This
object had no N or Sr enhancements.

\begin{figure}
\begin{center}
\includegraphics[scale=.70,angle=0]{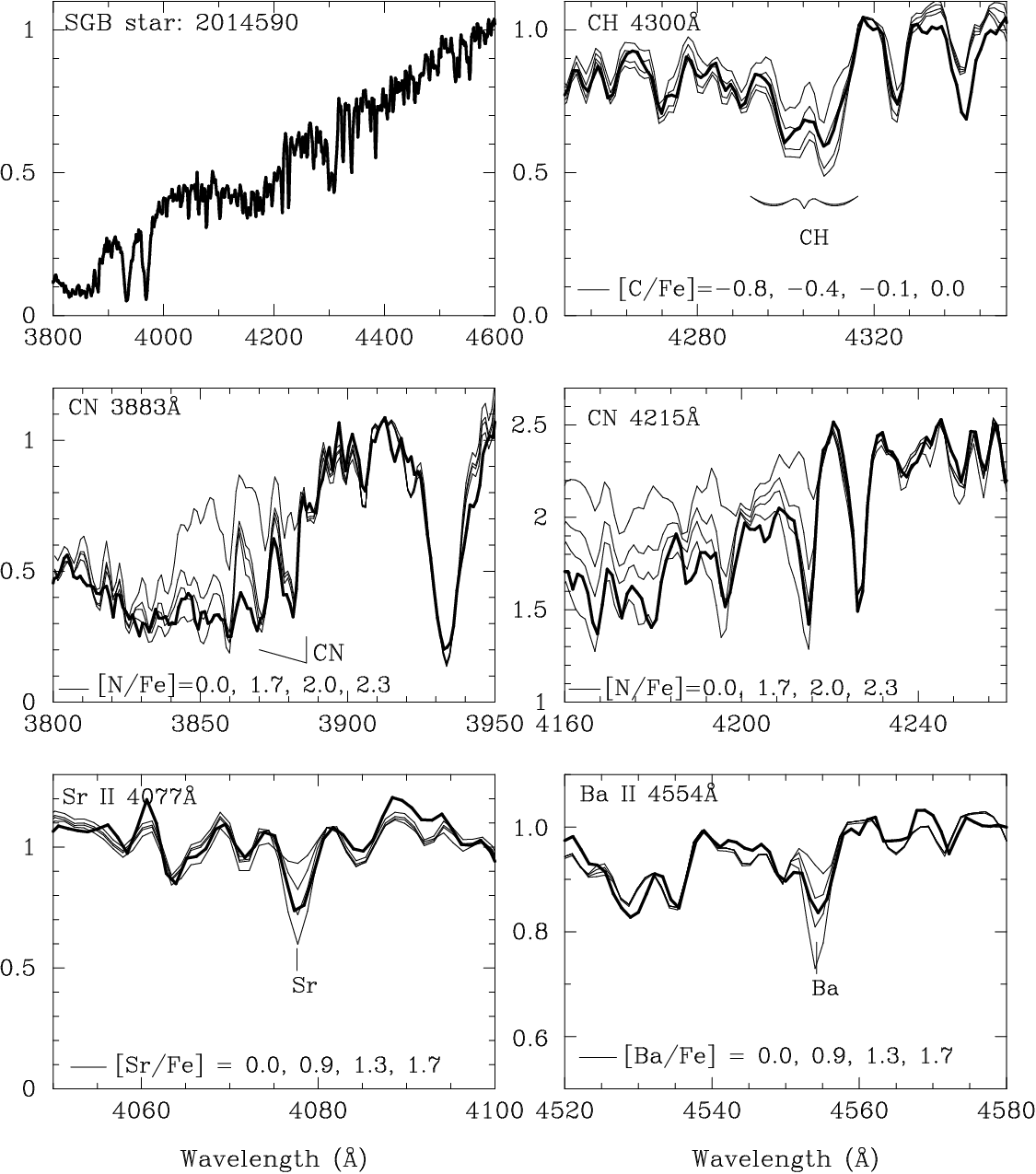}
\figcaption[f13.ps]{Observed and synthetic spectra of star 2014590
from the most metal-rich population, with [Fe/H]=--1.0~dex.  The top
left panel shows the spectrum of this object in the range
3800--4600{\AA}.  The top right panel shows the observed spectrum
(bold line) with synthetic spectra (light lines) in the region of the
G band.  The middle two panels show the CN features at $\sim$3883{\AA}
(left panel) and $\sim$4216{\AA} (right panel).  The bottom left panel
shows the Sr {\sc ii} 4077{\AA} line and the bottom right panel shows
Ba {\sc ii} 4554{\AA}. \label{pspec} }
\end{center}
\end{figure}

 Figure \ref{pspec} shows an example of a more typical member of the
3rd Pop, which has depleted carbon ([C/Fe]=--0.4). This was used to
obtain a N abundance from the CN features at 3883{\AA} and 4215{\AA},
yielding [N/Fe]=1.7 (middle panels).  This object also has enhanced Sr
and Ba, shown in the lower panels.  Another star in our sample,
3001426, has similar stellar parameters and abundances.

\section{Discussion} \label{disc_sect}

The abundance patterns found here, while somewhat limited, when taken
together with the data for the red giants, have the potential to
constrain putative sources of enrichment.  We begin with a short
discussion of possible sources of enrichment and follow with
constraints that the present results place on the enrichment
processes.

The origin of the abundance enhancements found here are most likely
primordial or the result of accretion events.  These abundance
patterns are unlikely to have been generated within the MS stars
themselves as the small convective region does not mix processed
materials from the stellar interior to its surface. Comparing these
results with those found on the RGB (\citealt{scl95},
\citetalias{nd95}) seems to suggest that we are finding similar
abundance enhancements in N and the light s-process elements, at least
on the SGB. However, most stars studied on the RGB are C depleted, or
near solar, and N is enhanced by $\sim$1.0~dex
(\citealt{cb86,pn89,bw93}; \citetalias{nd95}).  In RGB stars where the
C is depleted and N enhanced, the CN and ON cycles have been at work.
Whether this was in the stars themselves, in nearby objects whose
material was accreted, or occurred in objects from which these stars
formed is unclear.  However, primordial or accretion of C and/or O is
needed to explain the CO-strong stars that exist on the RGB
\citep{per80}. While the details of the cluster chemical evolution
remain in question, we discuss the various type of objects that might
have been responsible for the nucleosynthesis of the various elements
that are varying within the system.

\subsection{Possible Sources of Enrichment}

\subsubsection{Low Mass AGB stars}

Low mass (1--3 M$_{\odot}$) AGB stars undergo dredge-up episodes which
bring C and s-process elements into the convective region
\citep{gal98}.  Because of their low mass, these objects do not have
hot bottom burning (HBB) episodes which would produce N and reduce C
via the CNO cycle.  The amount of s-process elements that are produced
is dependent on neutron density and the availability of the seed Fe
nuclei \citep{smi05}. The neutron source is the
\mbox{$^{13}$C($\alpha$,n)$^{16}$O} reaction.  The s-process is
believed to occur in a $^{13}$C pocket which is produced at the third
dredge-up episode \citep{gal98,ll05}. Specifically, hydrogen downflow
from the envelope penetrates the $^{12}$C intershell region and at H
reignition a $^{13}$C enhanced zone is formed.  This $^{13}$C source
releases neutrons giving rise to efficient s-processing that is
dependent on the mass fractions of H, $^{4}$He, $^{13}$C, $^{14}$N and
metallicity.  For the C enhanced stars where we do not detect any Sr
enhancement (s-process), it may be that we simply do not have the
sensitivity to detect any enrichment.  Alternatively, if there was
little or no $^{13}$C pocket there was no source of neutrons for the
s-process to occur.  Thus, depending on the size and efficiency of the
$^{13}$C pocket, these AGB stars may account for the enhancements of C
with and without Sr found in our sample.

\subsubsection{Intermediate Mass AGB stars}

Intermediate mass (3--8 M$_{\odot}$) AGB stars are more likely to have
HBB, thereby producing a site for the operation of the CN cycle and
processing of C into N \citep{vdm02}.  Production of s-process
elements is not expected to be very high in these objects due to the
shorter duration of the $^{13}$C pocket \citep{lat04}.  Added to this
is lower neutron density from the $^{22}$Ne($\alpha$,n)$^{25}$Mg
reaction as the convective region becomes smaller in more massive
stars.  These stars may account for the objects with N enhancements
with little or no s-process element enhancements.

\subsubsection{Massive  Stars}

\citet{mm06} have recently presented models of rotating massive
($\sim$60 M$_{\odot}$) stars.  These objects produce stellar winds
with large excess of He and N, but no C or O excess.  Their products
are consistent with the abundances seen on the blue main sequence
(bMS) (see \citealt{pio05}), and may be responsible for some of the
enhanced objects found here.  Recently, abundance patterns of C, N, O,
Na and Li obtained from models of fast rotating, massive stars were
shown by \citet{dec06} to be similar the chemical anomalies observed
in normal globular cluster stars. Whether these two sets of models
also synthesize s-process elements is unclear but they may account for
objects with enhancements in N.

\citet{smi06} discussed the prospect of Wolf-Rayet and OB stars as
enrichment sources for globular cluster anomalies.  These objects (WN)
may produce large amounts of nitrogen, but a top-heavy stellar mass
function is required to generate the large amounts of N needed to
explain the observations in globular clusters.  A similar scenario is
needed in the \citet{mm06} models.

\subsection{Observational Constraints on the Enrichment of {\wcen}}

Similarities exist between the carbon-rich objects found on the MS and
those found on the RGB.  Here, two stars out of 195 ($\sim$1\%) were
found to have [C/Fe]$>$1.0 (these are the metal-poor stars 7007334
and 9005309). There is one possible subgiant CH star, 8001811, which
translates to $\sim$0.5\% of the total number of SGB stars with CH
star properties.  These percentages are similar to that found on the
RGB for stars brighter than V$\sim$13 in the Woolley et al. (1966)
catalogue, where there are four CH stars (55, 70, 279, 577, see
e.g. \citetalias{nd95}) out of a possible $\sim$600, equating to
0.7\%.  This suggests that one might expect to find an approximately
one percent incidence of CH stars in any sample at most stages of
evolution in the cluster (i.e. excluding blue horizontal branch
objects), and indicates that the CH stars are not limited to a
particular evolutionary phase.

It is interesting that the C-rich ([C/Fe]$\geq$0.5) stars occur
predominately at low metallicities and we find virtually none in the
3rd Pop.  This may result from small number statistics.  In the first
and second populations we find that 10\% of objects have enhanced C.
If the same percentage applies to the third population this would
equate to only two stars and it is not outside the realm of
possibility that they were simply not found in our search.  On the
other hand, it may be a result of the material from which these stars
formed being low in C to begin with.  The source of the carbon
enrichment could be low-mass AGB stars or perhaps SNe II.  In the
latter case, the outer layers escaped while the material close to the
iron core did not.  Low mass AGB stars also produce s-process
elements, depending on initial mass and metallicity. Therefore, they
are able to account for the range in s-process enhancement seen in
many of the C-rich stars.

Inspection of Figure \ref{abpat} and Table \ref{tbl-4} show there are
$\sim$10\% of the total sample (excluding the two metal-poor CH stars)
with [C/Fe]$\geq$0.5.  This is unusual when compared with the RGB
abundances in {\wcen}, as previous studies have not found any stars
(excluding CH stars) with carbon enhancements at these levels. Carbon
abundances of main sequence and subgiant branch stars in other
globular clusters also do not show any such enhancements.
\citet{car05} studied carbon, nitrogen and oxygen in dwarfs and
subgiant branch stars in NGC~6397, NGC~6752 and {\tuc}.  For dwarfs,
carbon abundances were less than [C/Fe]=0.50 for NGC~6397, [C/Fe]=0.2
for NGC~6752 and [C/Fe]=--0.1 for {\tuc}.  The subgiants showed
abundance levels of [C/Fe]$\leq$0.15 for NGC~6397, and
[C/Fe]$\leq$--0.10 for NGC~6752 and {\tuc}.

These C enhanced objects on the MSTO of {\wcen} may be the
evolutionary precursors to the CO-strong objects on the RGB
\citep{per80}.  Of the stars that are investigated in \citet{per80},
28 come from the unbiased sample of \citet{cs73}.  The CO-strong stars
represent 8 out of these 28 stars, or $\sim$30\% of RGB objects.  If
the 10\% of C-rich ([C/Fe]$\geq$0.5) stars found on the MSTO are the
evolutionary precursors, one would expect that more of these objects
should have been found.  However, our choice of the C-rich cutoff (at
[C/Fe]=0.5) is arbitrary and it is not known what the link is between
CO-strong status and carbon abundance on the MSTO.  If the C-rich
cutoff was lower we would have found a higher percentage of objects
that are C-rich on the MSTO.  It would be useful in future
investigations to determine carbon abundances for the whole sample to
confirm the connection between the C-enhanced objects at the MSTO and
the CO-strong objects on the RGB.

 The carbon abundances found here in the 3rd Pop. are low when
compared with those determined for dwarfs by \citet{car05} in
NGC~6397, NGC~6752 and {\tuc}. The lowest abundances found for
NGC~6752 and {\tuc} were [C/Fe]=--0.2 and [C/Fe]=--0.13, respectively.
The origin of these depletions for the most metal-rich group is not
well understood.  \citet{pan03} did not investigate carbon abundances
for the most metal-rich RGB sample in {\wcen}, and of the few
metal-rich RGB stars studied by \citetalias{nd95} carbon abundances
were found to be [C/Fe]$\sim$--0.3.  If evolutionary mixing on ascent
of the RGB occurred, one would expect even lower carbon abundances
than those found on the MS, but for the two metal-rich RGB stars
studied by \citetalias{nd95} this was not found to be the case.
Estimates of the level of carbon depletions for the 1st and 2nd
Pops. would be of great interest to be able to compare them not only
with those found in the 3rd Pop., but also with those on the RGB.
\citet{pio05} found [C/Fe]=0.0 for both the metal-poor and
metal-intermediate main sequences (rMS and bMS, respectively) using
composite spectra of stars belonging to both populations.

The nitrogen enhancements seen at the MSTO across all three
populations are quite substantial with [N/Fe] as large as 2.2.  These
abundances are slightly larger than those seen in other clusters
\citep{car05}. The nitrogen enhancements in NGC~6752 reach [N/Fe]=1.7
for dwarfs and [N/Fe]=1.3 for subgiants.  The maximum N enhancements
for {\tuc} are lower and only reach [N/Fe]=1.1 for the subgiants,
while our third population shows enhancements greater than [N/Fe]=1.1
for 37.5\% of the group.  \citet{pio05} investigated the two distinct
main sequences in {\wcen} and found abundances of C, N and Ba using
composite spectra of a number of stars in each sequence.  They found
[N/Fe]$\sim$1.0 for the blue main sequence (bMS), which is equivalent
in metallicity to the 2nd Pop. discussed here. This is lower by a
factor of 3--10 for the abundances of N found here.  This may be due
to a systematic offset in the abundances found here compared with
those found by \citet{pio05}.  On the {\wcen} RGB, \citetalias{nd95}
also found N abundances to be less than those found here by
$\sim$0.7~dex.

While Sr abundances have not been extensively investigated on the RGB,
lines of other light s-process elements such as Y and Zr have been
analyzed (\citetalias{nd95} \citealt{scl95, smi00}).  In general, it
was found that the abundance of the s-process elements relative to
iron increased as a function of metallicity.  There are no
correlations of s-process elemental abundances with other light
elements where variations are seen, such as C, N, O, Na and Al
\citepalias{nd95}.  Qualitatively, this is similar to what is found
here.  The incidence of Sr enhancement relative to iron increases as a
function of metallicity (see Table \ref{tbl-4}).  This fits in with
the idea that low to intermediate mass AGB stars contributed to the
gas from which the later generations of stars formed.  A comparison
between the Sr abundances found for RGB and MS stars will be discussed
further in a forthcoming paper (Stanford et al. 2007, in preparation).

It is not clear if intermediate mass AGB stars are responsible since
s-processing is not very efficient in these objects.  To account for
the N and Sr enhanced stars favorable conditions of N and s-process
element production in intermediate mass AGB stars are needed.
Alternatively, the enriched objects may be accounted for by several
different sources.  For instance, massive rotating stars may account
for the N overabundances while low-mass AGB may account for the
s-process enrichment.  This latter scenario is supported by the fact
that no strong correlation is seen between nitrogen and strontium
(s-process) abundances.

Other globular clusters, that do not show metallicity (i.e. iron)
spreads, show variations in the abundances of light elements to
varying degrees, as described in \S \ref{intro}.  Several clusters
show CN--CH bimodality not only on the RGB but also on the MS
\citep{can98, bri04, ss91, gra01a, coh99, bc01, rc02, cbs02, bcs04,
dac04}.  In this respect, the spread in C and N on the MSTO of {\wcen}
is not unique and these results demonstrate the variations are a
product of environment since field stars do not show similar patterns.
Variations in other light elements such as Na, O, Mg and Al are also
found on the MS in some clusters \citep{gra01b}.  These elements have
yet to be studied on the MS level in {\wcen} to determine if the same
effect is seen in this cluster.  The theoretical yields of AGB stars
at present fail to explain the abundance patterns seen in the light
elements in globular clusters (see e.g. \citealt{fen04}).  That is,
there is more to the enrichment scenario in other clusters that is yet
to be explained and this complication may also be the case for
{\wcen}.

\subsubsection{Constraints on the Role of Binarity}

If the enhancements are due to past accretion events as part of binary
systems, the fraction of objects with overabundances indicates {\wcen}
should have a large binary fraction, of order $\sim$20\%.  In a radial
velocity study of red giants in {\wcen} spanning over a decade,
\citet{may96} estimated the global binary frequency is 3--4\%.
Further, they reported that most of the giants with known chemical
peculiarities, such as Ba, CH and S stars, have constant velocities,
suggesting that they are not members of binary systems.  However,
there is the possibility that these objects may have been former
binaries that were disrupted by close encounters with nearby stars.
The same argument can be applied to the MS stars, where the fraction
of C, N and Sr enhanced stars exceeds the observed binary frequency
for giants in the cluster.

\section{Conclusion}

A low resolution spectroscopic abundance analysis of a sample of 392
{\wcen} members revealed objects enhanced in carbon, nitrogen and/or
strontium.  These enhanced objects were analyzed in detail and
abundances ratios of carbon, nitrogen and strontium relative to iron
determined.  The abundances revealed patterns that can possibly be
explained by low and intermediate mass AGB and/or rotating massive
star nucleosynthesis.  The enhancements are either primordial or from
accretion events, rather than evolutionary.  Higher resolution, and
higher S/N data are needed to further constrain the possible sources
of these abundances anomalies.  From such data, abundances of a large
number of elements such as the light elements (O, Na, Mg, Al), other
light and heavy s-process elements, and iron peak elements, will be
essential in obtaining another piece of the $\omega$ Cen puzzle.

\acknowledgements
We thank the Director and staff of the Anglo-Australian Observatory
for the use of their facilities.


\clearpage
\input{tab1.tex}

\input{tab2.tex}

\input{stub.tab3.tex}

\clearpage
\input{tab4.tex}

\input{tab5.tex}


\clearpage

\clearpage

\clearpage

\clearpage

\clearpage

\clearpage

\clearpage

\clearpage

\clearpage

\clearpage

\clearpage

\clearpage

\clearpage

\end{document}

%% file: tab1.tex
\begin{table}
\begin{center}
\tablenum{1}
\caption{Line Index Wavelength Bands ({\AA})
\label{tbl-1}}
\begin{tabular}{lccc}
\tableline\tableline
Line              & Line Band & Blue Sideband & Red Sideband \\
(1)               &  (2)      &  (3)          & (4)          \\
\tableline	      
CH -- G band      & 4292.5-4317.5  &  4247.0-4267.0  &  4357.0-4372.0  \\ 
Sr {\sc II} 4077  & 4074.7-4080.7  &  4048.0-4060.0  &  4085.0-4092.0  \\
Sr {\sc II} 4215  & 4212.0-4218.0  &  4201.0-4211.0  &  4245.0-4257.0  \\
Ba {\sc II} 4554  & 4551.0-4557.0  &  4505.0-4512.0  &  4573.0-4579.0  \\
\tableline
\tableline
\end{tabular}
\end{center}
\end{table}

%% file: tab2.tex
\begin{table}
\begin{center}
\tablenum{2}
\caption{Solar abundances and gf values used for s-process elements.
\label{tbl-2}}
\begin{tabular}{lccc}
\tableline\tableline
Line              & log(N/N$_{tot}$)$_{\odot}$ & gf value \\
(1)               &  (2)                       &  (3)     \\
\tableline	      
Sr {\sc II} 4077  & --9.10                     & 1.0460   \\ 
Sr {\sc II} 4215  & --9.10                     & 0.7161   \\
Ba {\sc II} 4554  & --9.85                     & 1.7050   \\
\tableline
\tableline
\end{tabular}
\end{center}
\end{table}

%% file: stub.tab3.tex
\begin{deluxetable}{cccccccccrrrr}
\tabletypesize{\scriptsize}
\tablecolumns{13}
\tablewidth{0pt}
\tablenum{3}
\tablecaption{Stellar parameters, Indices and Abundances for {\wcen} sample. The full version of this table appears in the electronic version \label{tbl-3}}
\tablehead{\colhead{ID}&\colhead{V mag}&\colhead{B--V}&\colhead{Teff}&\colhead{logg}&\colhead{[Fe/H]}&\colhead{G Band}&\colhead{S3839}&\colhead{Sr}&\colhead{Ba}&\colhead{[C/Fe]}&\colhead{[N/Fe]}&\colhead{[Sr/Fe]} \\ \colhead{}&\colhead{}&\colhead{}&\colhead{}&\colhead{}&\colhead{}&\colhead{}&\colhead{}&\colhead{4077\AA}&\colhead{4554\AA}&\colhead{}&\colhead{}&\colhead{} \\ \colhead{(1)}&\colhead{(2)}&\colhead{(3)}&\colhead{(4)}&\colhead{(5)}&\colhead{(6)}&\colhead{(7)}&\colhead{(8)}&\colhead{(9)}&\colhead{(10)}&\colhead{(11)}&\colhead{(12)}&\colhead{(13)} 
}
\startdata
 1000812 & 17.470 & 0.610 &  5874 &  3.8 & --1.64 &  1.42 & --0.06 &  0.19 &  0.09 & \nodata & \nodata & \nodata \\
 1002064 & 18.290 & 0.570 &  6066 &  4.3 & --1.41 &  2.42 & --0.06 &  0.28 &  0.09 & \nodata & \nodata & \nodata \\
 1002884 & 17.490 & 0.600 &  5901 &  3.8 & --1.76 &  1.51 & --0.11 &  0.11 &  0.19 & \nodata & \nodata & \nodata \\
 1004374 & 17.500 & 0.620 &  5883 &  3.8 & --1.72 &  2.71 & --0.14 &  0.18 &  0.20 &  0.50 &  0.00 &  0.00 \\
 1005088 & 17.390 & 0.730 &  5443 &  3.6 & --1.58 &  5.79 & --0.06 &  0.18 &  0.29 & \nodata & \nodata & \nodata \\
 1005996 & 17.320 & 0.720 &  5572 &  3.7 & --1.49 &  4.95 & --0.10 &  0.37 &  0.24 & \nodata & \nodata & \nodata \\
 1006065 & 17.510 & 0.650 &  5761 &  3.8 & --1.62 &  2.48 & --0.01 &  0.11 &  0.04 & \nodata & \nodata & \nodata \\
 1006286 & 17.380 & 0.710 &  5564 &  3.7 & --1.49 &  4.64 & --0.04 &  0.29 &  0.28 & \nodata & \nodata & \nodata \\
 1006625 & 17.540 & 0.610 &  5869 &  3.8 & --1.78 &  2.07 & --0.11 &  0.06 &  0.19 & \nodata & \nodata & \nodata \\
 1006759 & 18.120 & 0.510 &  6282 &  4.2 & --1.71 &  0.73 & --0.10 &  0.07 &  0.22 & \nodata & \nodata & \nodata \\
 1006812 & 17.430 & 0.640 &  5772 &  3.8 & --1.49 &  2.71 & --0.07 &  0.07 &  0.25 & \nodata & \nodata & \nodata \\
 1006962 & 17.870 & 0.670 &  5672 &  3.9 & --1.59 &  4.43 & --0.11 &  0.41 &  0.29 &  0.40 &  0.50 &  1.40 \\
 1007266 & 18.430 & 0.530 &  6179 &  4.3 & --1.50 &  0.74 & --0.13 & --0.09 &  0.08 & \nodata & \nodata & \nodata \\
 1007362 & 17.450 & 0.680 &  5681 &  3.7 & --1.65 &  4.29 & --0.10 &  0.29 &  0.12 &  0.50 &  0.00 &  0.00 \\
 1007414 & 17.840 & 0.630 &  5869 &  3.9 & --1.44 &  2.89 & --0.13 &  0.25 &  0.27 & \nodata & \nodata & \nodata \\
\enddata 
\end{deluxetable}

%% file: tab4.tex
\begin{deluxetable}{lrrrrrrrrrrrrrr}
\tablecolumns{15}
\tabletypesize{\footnotesize}
\tablewidth{0pc}
\tablenum{4}
\tablecaption{Enhancement of C, N and Sr statistics. \label{tbl-4}}
\tablehead{
\colhead{Sample }   & \multicolumn{2}{c}{Total \#} & \colhead{} & \multicolumn{2}{c}{Normal} & \colhead{} & \multicolumn{2}{c}{[C/Fe]$\geq$0.5} & \colhead{} & \multicolumn{2}{c}{[N/Fe]$\geq$1.0} & \colhead{} & \multicolumn{2}{c}{[Sr/Fe]$\geq$1.0} \\ 
\cline{2-3}
\cline{5-6}
\cline{8-9} 
\cline{11-12} 
\cline{14-15} \\
\colhead{}         & \colhead{MS} &  \colhead{SGB} &  \colhead{} &  \colhead{MS} &  \colhead{SGB} &  \colhead{} &  \colhead{MS} &  \colhead{SGB} &  \colhead{} &  \colhead{MS} &  \colhead{SGB} &  \colhead{} &  \colhead{MS} &  \colhead{SGB}  
}
\startdata
Full Sample                     & 195  & 197   &  &  88.2\%  & 65.0\% &   & 4.1\%  & 16.8\% &   & 6.2\%  & 16.2\% &   & 5.1\%  & 14.2\%     \\
1st Population\tablenotemark{1} & 116  & 118   &  &  97.4\%  & 75.4\% &   & 1.7\%  & 14.4\% &   & 0.0\%  &  8.5\% &   & 0.9\%  & 9.3\%      \\
2nd Population\tablenotemark{1} &  61  &  73   &  &  80.3\%  & 49.3\% &   & 8.2\%  & 21.9\% &   & 9.8\%  & 26.0\% &   & 8.2\%  & 19.2\%     \\ 
3rd Population\tablenotemark{1} &  18  &   6   &  &  55.6\%  & 50.0\% &   & 5.6\%  &  0.0\% &   &33.3\%  & 50.0\% &   &22.2\%  & 50.0\%       \\
\enddata
\tablenotetext{1}{1st Population: [Fe/H]$<$--1.5; 2nd Population: --1.5$\leq$[Fe/H]$<$--1.1; 3rd Population: [Fe/H]$\geq$--1.1}
\end{deluxetable}

%% file: tab5.tex
\begin{deluxetable}{lrrrrrrrrr}
\tablecolumns{10}
\tabletypesize{\scriptsize}
\tablewidth{0pc}
\tablenum{5}
\tablecaption{Abundance Patterns found on the Main Sequence of $\omega$ Cen. \label{tbl-5}}
\tablehead{
\colhead{Abundance Pattern\tablenotemark{1}}    & \colhead{} & \multicolumn{2}{l}{1st Pop.\tablenotemark{2}} & \colhead{} & \multicolumn{2}{l}{2nd Pop.\tablenotemark{2}} & \colhead{}  & \multicolumn{2}{l}{3rd Pop.\tablenotemark{2}}\\ 
\colhead{}    & \colhead{} & \colhead{MS}   & \colhead{SGB}   & \colhead{} & \colhead{MS}   & \colhead{SGB}   & \colhead{} & \colhead{MS}   & \colhead{SGB}   
}
\startdata
C only enh.          &  & 2      & 13      &   & 3     & 10    &   & 1      & 0                    \\
C + N  enh.          &  & 0      & 0       &   & 0     & 0     &   & 0      & 0                    \\
C + Sr enh           &  & 0      & 3       &   & 2     & 6     &   & 0      & 0                    \\
C + N + Sr enh.      &  & 0      & 1       &   & 0     & 0     &   & 0      & 0                    \\
\cline{1-1}	       
\cline{3-4}	       
\cline{6-7}	       
\cline{9-10}	       
Total C enh.         &  & 2      & 17      &   & 5     & 16    &   & 1      & 0                    \\
\cline{1-1}	       
\cline{3-4}	       
\cline{6-7}	       
\cline{9-10}	       
                     &  &        &         &   &       &       &   &        &                      \\
N only enh.          &  & 0      & 5       &   & 4     & 13    &   & 3      & 0                    \\
N + C  enh.          &  & 0      & 0       &   & 0     & 0     &   & 0      & 0                    \\
N + Sr  enh.         &  & 0      & 4       &   & 2     & 6     &   & 3      & 3                    \\
N + C + Sr enh.      &  & 0      & 1       &   & 0     & 0     &   & 0      & 0                    \\
\cline{1-1}	       
\cline{3-4}	       
\cline{6-7}	       
\cline{9-10}	       
Total N enh.         &  & 0      & 10      &   & 6     & 18    &   & 6      & 3                    \\
\cline{1-1}	       
\cline{3-4}	       
\cline{6-7}	       
\cline{9-10}	       
                     &  &        &         &   &       &       &   &        &                      \\
Sr only enh.         &  & 1      & 3       &   & 1     & 2     &   & 3      & 0                    \\
Sr + C enh.          &  & 0      & 3       &   & 2     & 6     &   & 0      & 0                    \\
Sr + N enh.          &  & 0      & 4       &   & 2     & 6     &   & 3      & 3                    \\
Sr + C + N enh.      &  & 0      & 1       &   & 0     & 0     &   & 0      & 0                    \\
\cline{1-1}	       
\cline{3-4}	       
\cline{6-7}	       
\cline{9-10}	       
Total Sr enh.        &  & 1      & 11      &   & 5     & 14    &   & 6      & 3                    \\
\cline{1-1}	       
\cline{3-4}	       
\cline{6-7}	       
\cline{9-10}	       
                     &  &        &         &   &       &       &   &        &                      \\
C depl. only         &  & 0      & 0       &   & 0     & 0     &   & 4      & 0                    \\
C depl. + N enh.     &  & 0      & 0       &   & 0     & 0     &   & 3      & 0                    \\
C depl. + Sr enh.    &  & 0      & 0       &   & 0     & 0     &   & 1      & 0                    \\
C depl. + N + Sr enh.&  & 0      & 0       &   & 0     & 0     &   & 1      & 2                    \\
\cline{1-1}	       
\cline{3-4}	       
\cline{6-7}	       
\cline{9-10}	       
Total C dep.         &  & 0      & 0       &   & 0     & 0     &   & 9      & 2                    \\
\enddata
\tablenotetext{1}{Carbon enhanced: [C/Fe]$\geq$0.5; Nitrogen enhanced: [N/Fe]$\geq$1.0; Strontium enhanced: [Sr/Fe]$\geq$1.0; Carbon depleted: [C/Fe]$\leq$--0.2}
\tablenotetext{2}{1st Pop.: [Fe/H]$<$--1.5; 2nd Pop.: --1.5$\leq$[Fe/H]$<$--1.1; 3rd Pop.: [Fe/H]$\geq$--1.1}
\end{deluxetable}